\documentclass{JHEP3} 

\JHEPspecialurl{http://jhep.sissa.it/JOURNAL/JHEP3.tar.gz}

\usepackage{epsfig,multicol,amsmath}

\def\beq{\begin{equation}}
\def\eeq{\end{equation}}
\def\bea{\begin{eqnarray}}
\def\eea{\end{eqnarray}}

\def\eq#1{{Eq.~(\ref{#1})}}
\def\fig#1{{Fig.~\ref{#1}}}
\def\sec#1{{section~\ref{#1}}}
\def\rho{x}
\newcommand{\bas}{\bar{\alpha}_s}
\newcommand{\as}{\alpha_s}

\newcommand{\Lb}{\left(}
\newcommand{\Rb}{\right)}

\parskip=\itemsep               

\newcommand{\D}{\partial}
\newcommand{\h}{\frac{1}{2}}

\newcommand{\x}{\vec{x}}

\renewcommand{\theequation}{\thesection.\arabic{equation}}

\def\ap#1#2#3{     {\it Ann. Phys. (NY) }{\bf #1} (19#2) #3}
\def\arnps#1#2#3{  {\it Ann. Rev. Nucl. Part. Sci. }{\bf #1} (19#2) #3}
\def\npb#1#2#3{    {\it Nucl. Phys. }{\bf B#1} (19#2) #3}
\def\plb#1#2#3{    {\it Phys. Lett. }{\bf B#1} (19#2) #3}
\def\prd#1#2#3{    {\it Phys. Rev. }{\bf D#1} (19#2) #3}
\def\prep#1#2#3{   {\it Phys. Rep. }{\bf #1} (19#2) #3}
\def\prl#1#2#3{    {\it Phys. Rev. Lett. }{\bf #1} (19#2) #3}
\def\ptp#1#2#3{    {\it Prog. Theor. Phys. }{\bf #1} (19#2) #3}
\def\rmp#1#2#3{    {\it Rev. Mod. Phys. }{\bf #1} (19#2) #3}
\def\zpc#1#2#3{    {\it Z. Phys. }{\bf C#1} (19#2) #3}
\def\mpla#1#2#3{   {\it Mod. Phys. Lett. }{\bf A#1} (19#2) #3}
\def\nc#1#2#3{     {\it Nuovo Cim. }{\bf #1} (19#2) #3}
\def\yf#1#2#3{     {\it Yad. Fiz. }{\bf #1} (19#2) #3}
\def\cpc#1#2#3{    {\it Comp. Phys. Commun. }{\bf #1} (19#2) #3}
\def\dis#1#2{      {\it Dissertation, }{\sf #1 } 19#2}
\def\dip#1#2#3{    {\it Diplomarbeit, }{\sf #1 #2} 19#3 }
\def\ib#1#2#3{     {\it ibid. }{\bf #1} (19#2) #3}
\def\jpg#1#2#3{        {\it J. Phys}. {\bf G#1}#2#3}
\def\thefootnote{\fnsymbol{footnote}}

\title{\LARGE \bf Survival probability for  diffractive Higgs production in high density
QCD}
\author{\large  J.~Miller\thanks{Email:
jeremymi@post.tau.ac.il;}\,\, \\
Department of Particle Physics, School of Physics and Astronomy\\
Raymond and Beverly Sackler
 Faculty
of Exact Science\\  Tel Aviv University, Tel Aviv, 69978, Israel}

\abstract{ In this paper, the contribution of hard processes
described by the BFKL pomeron exchange, is taken into account by
calculating the first enhanced diagram. The survival probability is
estimated, using the ratio of the first enhanced diagram and the
single pomeron amplitude, taking into account all essential pomeron
loop diagrams in the toy model of Mueller. The triple pomeron vertex
is calculated explicitly in the momentum representation. This
calculation is used for estimating the survival probability,  It
turns out that the survival probability is small, at $0.4\%{}$. Hard
pomeron re-scattering processes contribute substantially to the
survival probability.}


\preprint{  TAUP -2???-07\\
hep-ph/0610427\\
\today}

\begin{document}

\def\thefootnote{\arabic{footnote}}
\section{Introduction}
\label{sec:Int}
\subsection{The survival probability}
\label{sec:sp3} The goal of this paper is to calculate the survival
probability, taking into account the contribution of hard processes
described by the BFKL pomeron exchange. The diffractive Higgs
production is a typical hard process, in which the Higgs is produced
from the one parton shower due to gluon fusion.
This process can be calculated in perturbative QCD.\\

The signature of this process is the existence of so called large
rapidity gaps (LRG), in which no particles are produced (see
Refs.\cite{1,2}). For the LHC energies and for diffractive Higgs
production at the c.m rapidity equal to zero, there are two rapidity
gaps. The first is between the right moving final protons and the
Higgs boson, the second is between the left fast moving proton and
the Higgs boson.\\

As was noticed by Bjorken \cite{2}, in hadron hadron collisions,
there is a considerable probability that more than one parton shower
can be produced. Therefore, one needs to suppress such a multi
parton shower production, since it can produce particles that fill
up the rapidity gap. This suppression can be characterized
by the survival probability \cite{2,3}.\\

To illustrate what survival probability is, it is instructive to
calculate it in the simple eikonal model for soft pomerons. Soft
pomeron means that there are no perturbative contributions from
short distances, and only soft non - perturbative processes
contribute to high energy asymptotic behavior. The survival
probability is defined in the eikonal formalism as \cite{2,3}
\begin{equation}
<{}|S^{2}|{}>{}=\frac{\int{}|\mathcal{M}\left(s,b\right)|^{2}e^{-\Omega(b)}d^{2}b}{\int{}|\mathcal{M}\left(s,b\right)|^{2}d^{2}b}\label{E:1.1}
\end{equation}
where $\mathcal{M}$ is the amplitude for the hard process under
consideration, in impact parameter space (where $b$ is the impact
parameter), at the centre of mass energy $\sqrt{s}$. In
this paper, this is the amplitude of diffractive Higgs production
from one parton shower. $e^{-\Omega{}\left(b\right)}$ gives the
probability that additional inelastic scattering will not occur
between the two partons at impact parameter $b$.
$\Omega\left(b\right)$ is called the opacity or optical density.
Therefore, the numerator is the amplitude for
the exclusive process, while the denominator is the same process, due to the exchange of one pomeron.\\

The survival probability was estimated in Ref.\cite{6}, in the
eikonal approach for exclusive central diffractive production at the
LHC. The survival probability here was given for the process
illustrated in \fig{lrg}, in terms of the impact parameter $b$.
Generally, in all these models the survival probability is given by
the expression
\begin{equation}
<|S^{2}|>=\frac{\int{}d^{2}b_{1}d^{2}b_{2}\left(A_{H}(b_{1})A_{H}(b_{2})\Lb\,1-A_{s}(\left(b_{1}+b_{2}\right)\Rb\,^{2})\right)^{2}}{\int{}d^{2}b_{1}d^{2}b_{2}\left(A_{H}(b_{1})A_{H}(b_{2})\right)^{2}}\label{E:1.2}
\end{equation}
\FIGURE[ht]{ \centerline{\epsfig{file=
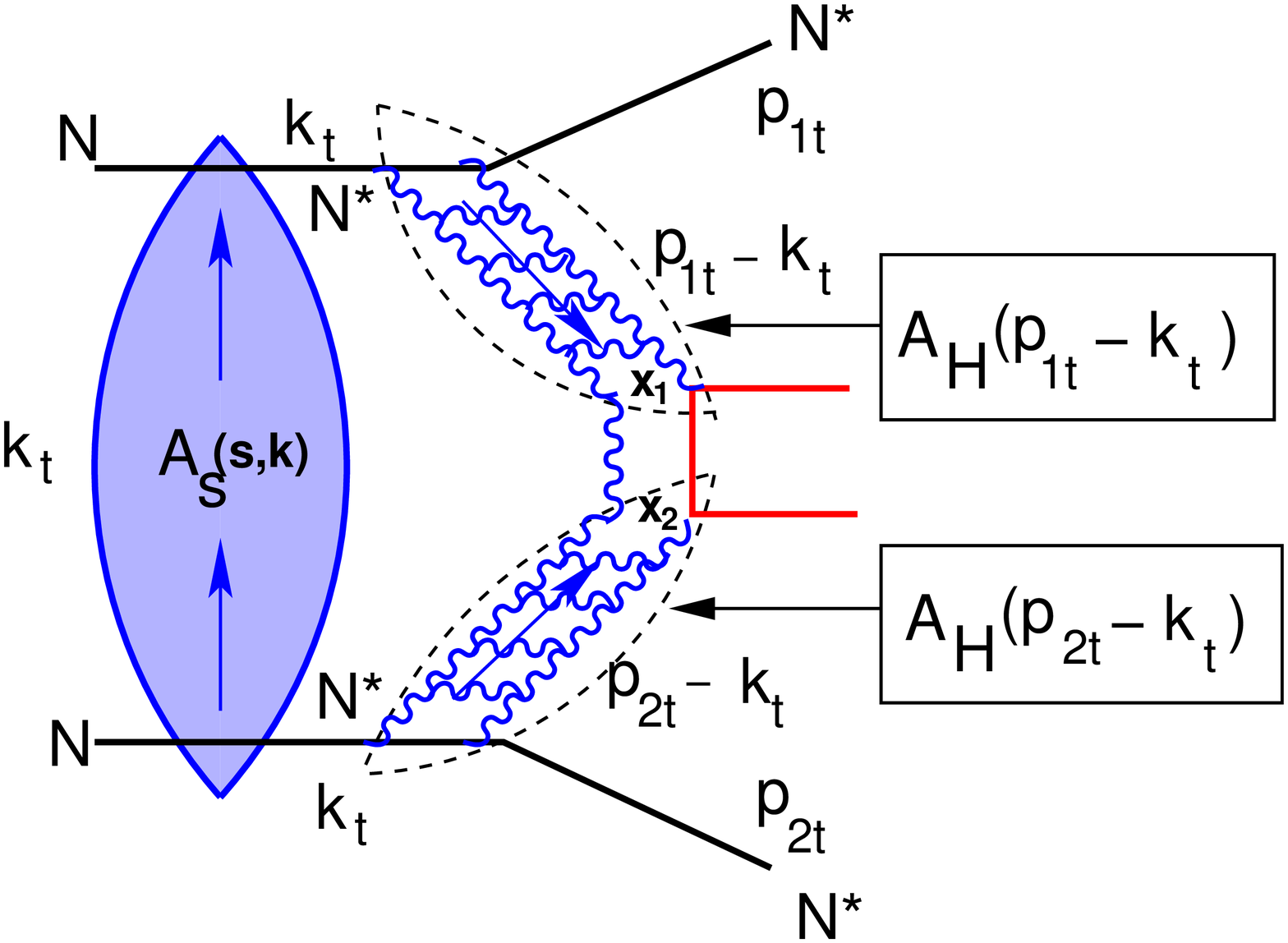,width=100mm,height=75mm}} \caption{ Central diffractive
production in the two channel eikonal model in proton scattering due
to pomeron exchange. } \label{lrg}} $A_{H}(b)$ is the hard pomeron
amplitude in impact parameter space $b$ shown in \fig{lrg}. The
amplitude $A_{H}\left(b\right)$ for hard pomeron exchange can be
calculated in perturbative QCD, and is responsible for the
production of two gluon jets, with BFKL ladder gluons between them
(see \fig{lrg}). In this model \cite{6}, the hard pomeron in
\fig{lrg} emits the Higgs. $A_{H}(b)$ is given in the impact
parameter $b$ representation  by the expression \cite{6}
\begin{equation}
A_{H}(b)=\frac{1}{\pi{}R_{H}^{2}}e^{-\frac{b^{2}}{R_{H}^{2}}}\label{E:1.3}
\end{equation}
$R_{H}^{2}=7.2GeV^{-2}$. As shown in \fig{lrg}, $A_{H}(b_{1})$ and
$A_{H}(b_{2})$ denote the hard pomeron amplitude above and below the
Higgs signal respectively. The contribution $A_{s}$, shown in
\fig{lrg}, denotes the soft pomeron amplitude. $\Lb\,1-A_{s}\Rb\,$
includes all possible initial state interactions due to the exchange
and interaction of soft Pomerons. $\Lb\,1-A_{s}\Rb\,$ also includes
the possibility that the two initial nucleons in \fig{lrg}, do not
interact at all.\\

The survival probability was found to be 5 - 6 $\%$ for the single
channel model. In the two channel model, the survival probability
here is $2.7\%{}$ at the LHC energy of $\sqrt{s}=14000$ $GeV$. The
upper bound for the survival probability, in the constituent quark
model (CQM), was found to be $6.0\%{} \pm{}0.1\%{}$ at the LHC
energy. This is almost the same as the survival probability found in
the single channel model. The two upper bounds, intercept at an
energy just above the typical LHC energy. This suggests, that the
upper bound for the survival
probability, should be $2\%{}-3\%{}$ for measurements at the LHC.\\

The first attempt to estimate the contribution of hard (semi - hard)
processes to the value of the survival probability, was made by
Bartels, Bondarenko, Kutak and Motyka in \cite{7}. They considered
the contribution of this "fan" pomeron diagram, to the value of the
survival probability, and  found that this contribution is rather
large. Namely, the value ranges from $3.17\%\,$ for $\as=0.15$, to
$1.6\%$ for $\as=0.25$, (where $\as$ is the QCD coupling).\\

The aim of this paper, is to calculate the BFKL pomeron (see
\fig{BFKL}), and the first enhanced diagram for the BFKL pomeron
(see \fig{BFKLenh}). These calculations are in the symmetric QCD
dipole approach. Because in proton proton scattering, there
is no reason to assume that the mean field approximation, based on
the "fan" diagram, can work. The ratio of the two contributions of
\fig{BFKL} and \fig{BFKLenh} are calculated, and used to
estimate the value of the survival probability.\\

This paper is organised in the following way.
In \sec{sec:BFKL1}, the coupling of the BFKL pomeron to the colour
dipole (\sec{sec:vert}), and the triple pomeron vertex
(\sec{sec:3p}), are calculated in the momentum representation. Using
these results, the BFKL pomeron amplitude shown in \fig{BFKL} is
calculated (\sec{sec:BFKL}), and the first enhanced diagram shown in
\fig{BFKLenh}, is calculated (\sec{sec:BFKLenh}).\\

Section \ref{sec:sp} is devoted to the survival probability,
estimated in the QCD dipole approach. The ratio of the two
contributions of \fig{BFKL} and \fig{BFKLenh} is calculated, which
is used to estimate the value of the survival probability
(\sec{sec:sp2}). It turns out, that this ratio is not small, and
indicates the importance of taking into account all enhanced
diagrams. Therefore, in \sec{sec:toy}, all enhanced diagrams are
summed in the toy model (see Ref.\cite{toy}). The fact that
the two dipoles have different sizes is neglected. From the
calculation of the ratio of \fig{BFKL} and \fig{BFKLenh}, the value
of the parameter $d$ of this model is determined, (d is the low
energy amplitude for one pomeron exchange). Using this parameter,
the value of the survival probability was estimated as the ratio of
the diffractive Higgs production in this model, and Higgs production
in one parton shower (for single pomeron exchange).
It turns out that the survival probability is rather small.\\

In the conclusion, the results for the value of the survival
probability is presented. A discussion is given on the dependence of
the value of the survival probability, on the choice of intercept of
the BFKL pomeron. The significance of higher order hard rescattering
contributions to the survival probability, is also discussed.

\section{The conformal eigenfunctions of the vertex operator and
the triple pomeron vertex} \numberwithin{equation}{subsection}
\label{sec:BFKL1}

All calculations are carried out in the momentum representation, and
 the strategy and notation of Ref. \cite{31b} is closely
followed. Firstly, the pomeron coupling to the QCD colour dipole is
introduced (see \fig{vpomq}) in the momentum representation.
Secondly, an explicit expression for the triple pomeron vertex (see
\fig{vpomqk}) is derived. Using both these formulae, the BFKL
pomeron (\fig{BFKL}), and the first enhanced diagram
(\fig{BFKLenh}), are calculated in the symmetric QCD dipole
approach.

\subsection{The BFKL pomeron vertex function}
\label{sec:vert}

\FIGURE[ht]{\begin{minipage}{35mm}
\centerline{\epsfig{file=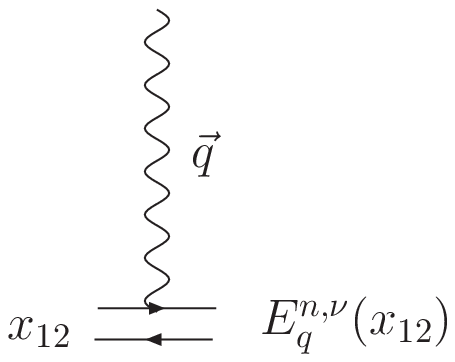,width=30mm,height=25mm}}
\end{minipage}
\caption{ The vertex of interaction of Pomeron with the dipole. }
\label{vpomq} }

The vertex coupling the BFKL pomeron to the couple dipole is
illustrated in \fig{vpomq}. Here, $\vec{q}$ is the momentum
transferred along the pomeron, and $\x_{12}$ is the transverse size
of the dipole. In the notation of Refs. \cite{11,16} , the
eigenfunctions for the vertex in coordinate space are defined as

\beq
E^{n,\nu}\Lb\,x_{01},x_{20}\Rb=\Lb-1\Rb^n\Lb\frac{x_{10}x_{20}}{x_{12}}\Rb^{\gamma-\frac{1}{2}}\Lb\frac{\bar{x}_{10}\bar{x}_{20}}{\bar{x}_{12}}\Rb^{\bar{\gamma}-\frac{1}{2}}\label{E:c1}
\eeq

where $x_{ij}=x_{i}-x_{j}$ and $x_{i}$ are the transverse
coordinates. The conformal dimensions are defined as

\begin{align} \gamma=\frac{n}{2}-i\nu& &\bar{\gamma}=-\frac{n}{2}-i\nu \label{E:c2}
\end{align}

$n$ is the conformal spin, and is an integer. The energy levels of
the pomeron are the BFKL eigenvalues given by \cite{11} \beq
\label{OM}
\omega(n,\nu)\,=\bas\,\chi\Lb\,\gamma\Rb\,=\bas\,\Lb\,2\psi\Lb\,1\Rb\,-\psi\Lb\,\gamma\Rb\,-\psi\Lb\,1-\gamma\Rb\,\Rb\,\eeq

where in this paper the notation

\beq
\bas\,=\,\frac{\as\,N_c}{\pi}\label{astrong}
\eeq

is used, and where $\psi(f) = d \ln \Gamma(f)/d f$ and $\Gamma(f)$ is the Euler
gamma function. Since the only intercept $\omega(n=0,\nu)$ is
positive at high energies, the contribution with $n \neq 0$ can be
neglected. Lipatov in Ref. \cite{11} introduces the following mixed
representation of the vertex.

\beq
E^{n,\nu}_{q}\Lb\x\Rb=\frac{2\pi^2}{b_{n,\nu}}\frac{1}{\vert\x\vert}\int{}d^{2}Re^{i\vec{q}\cdot\,\vec{R}}E^{n,\nu}\Lb\,R+\frac{\rho}{2},R-\frac{\rho}{2}\Rb\label{E:cc3}
\eeq

where \cite{11}

\beq
b_{n,\nu}=\frac{2^{4i\nu}\pi^3}{\frac{n}{2}-i\nu}\frac{\Gamma\Lb\frac{n}{2}-i\nu+\frac{1}{2}\Rb\Gamma\Lb\frac{n}{2}+i\nu\Rb}{\Gamma\Lb\frac{n}{2}+i\nu+\frac{1}{2}\Rb\Gamma\Lb\frac{n}{2}-i\nu\Rb}\label{E:c4}
\eeq

A more convenient expression of \eq{E:cc3} for the vertex was
calculated in Ref. \cite{16} and is given as

\bea
E^{n,\nu}_q\Lb x_{12} \Rb\,\,=&&\,\Lb\,qq^{\ast}\Rb\,^{i\nu}\,2^{-6\,i\,\nu}\,\Gamma^2(1 - i\nu)\,\times \label{EQNA}\\
&& \,\Lb J_\gamma\Lb \frac{q^*\,x_{12}}{4}\Rb\,J_{\tilde{\gamma}}\Lb
\frac{q\,x^*_{12}}{4}\Rb\,\,-\,\Lb\,-1\Rb\,^{n}\, J_{-\gamma}\Lb
\frac{q^*\,x_{12}}{4}\Rb\,J_{-\tilde{\gamma}}\Lb
\frac{q\,x^*_{12}}{4}\Rb \Rb \nonumber \eea

where $J_{\gamma}$ are the Bessel functions of the first kind. In
\eq{EQNA}, $q$ and $q^{\ast}$ are the components of the momentum
$\vec{q}$ transferred along the pomeron, in the complex
representation. That is

\bea \vert\,q\vert\,&=&q\,q^{\ast}\nonumber\\
\mbox{where}\,\,\,\,\,\,\,\,\,\,\,q&=&q_x+iq_y\,\,\,\,\,\,\,\,\,\,\,\,\,q^{\ast}\,=\,q_x-iq_y\label{comp}\eea
 In
order to work in the momentum representation when calculating the
single pomeron amplitude (\fig{BFKL}), and the first enhanced
diagram (\fig{BFKLenh}), it is necessary to express the vertex
function explicitly in the momentum representation. In
Ref.\cite{31b} it was shown that in the momentum representation, the
vertex function is given by the following Fourier transform

\begin{equation}
E\Lb
\vec{p},\vec{q};\gamma\Rb\,=\frac{b_{n,\nu}}{2\pi^2}\int{}\frac{dx{}}{\sqrt{x{}}}\exp{}\left(-\frac{ip^{\ast}x{}}{2}\right)\int{}\frac{dx^{\ast}}{\sqrt{x^{\ast}}}\exp{}\left(-\frac{ix^{\ast}p}{2}\right)E^{n,\nu}_{q}\left(x{}\right)
\label{E:c5}
\end{equation}
Here $\vec{p}$ denotes the momentum which is the conjugate variable
of the dipole size $x_{12}$. The complex representation, to express
the vector $\vec{p}$ in terms of its complex components $p$ and
$p^{\ast}$ (see \eq{comp}), is used in \eq{E:c5}. In Ref.
\cite{31b}, this integral is written in the following factorised
form

\bea E\Lb
\vec{p},\vec{q};\gamma\,\Rb&=&\frac{b_{n,\nu}}{2\pi^2}(q^2)^{- i
\nu}\,2^{-6\,i\,\nu}\,\Gamma^2(1 -
i\nu)\,\Lb\,\tilde{E}(p,q;\tilde{\gamma})\tilde{E}(p^{*},q^{*};\gamma)-\tilde{E}(p,q;-\tilde{\gamma})\tilde{E}(p^{*},q^{*};-\gamma)\Rb\label{ETIL}
\eea

where

\bea
\tilde{E}(p^*,q^*;\gamma)=\int\frac{dx}{\sqrt{x}}J_{\tilde{\gamma}}\Lb\frac{q^*x}{4}\Rb\,e^{-\frac{i}{2}p^*x}&\,\,\,\,\,\,\,
&\tilde{E}(p,q;\tilde{\gamma})=\int\frac{dx^*}{\sqrt{x^*}}J_{\gamma}\Lb\frac{qx^*}{4}\Rb\,e^{-\frac{i}{2}px^*}\label{ETIL1}
\eea

At this point, it is assumed that $n=0$, and hence
$\gamma=\tilde{\gamma}=-i\nu$ (see \eq{E:c2}). This is because the
only intercept $\omega\Lb\,n=0,\nu\Rb\,$ is positive at high
energies (see \eq{OM}), so the contribution $n\neq\,0$ is neglected
from now onwards. Let $\tilde{E}(p^*,q^*;\gamma)$ and
$\tilde{E}(p,q;\tilde{\gamma})$ be denoted as
$\tilde{E}(p^*,q^*;\nu)$ and $\tilde{E}(p,q;\nu)$ for $n=0$. After
integration over $x$ and $x^{\ast}$, the expressions for
$\tilde{E}\Lb\,q,p;\nu\Rb\,$ and
$\tilde{E}\Lb\,q^{\ast},p^{\ast};\nu\Rb\,$ are found to be
\cite{31b}

\bea\tilde{E}(p,q;\nu)&=&\Lb\,\frac{q}{8}\Rb^{-i\nu}\Lb\,-1\Rb^{-i\nu}\,i^{i\nu+\frac{1}{2}}2^{\frac{3}{2}-i\nu}p^{i\nu-\frac{1}{2}}\frac{\Gamma\Lb\,\frac{1}{2}-i\nu\Rb}{\Gamma\Lb\,1-i\nu\Rb\,}\,\,\,\notag\\
&\times\,&\,\,\,\,_2F_1\Lb\,\frac{1}{4}-\frac{1}{2}i\nu\,,\frac{3}{4}-\frac{1}{2}i\nu\,,1-i\nu\,,\frac{q^2}{4p^2}\Rb\,\notag\\
 \tilde{E}(p^*,q^*;\nu)&=&\Lb\,\frac{q^{\ast}}{8}\Rb^{-i\nu}\Lb\,-1\Rb^{-i\nu}\,i^{i\nu+\frac{1}{2}}2^{\frac{3}{2}-i\nu}\Lb\,p^{\ast}\Rb\,^{i\nu-\frac{1}{2}}\frac{\Gamma\Lb\,\frac{1}{2}-i\nu\Rb}{\Gamma\Lb\,1-i\nu\Rb\,}\,\,\,\notag\\
&\times\,&\,\,\,\,_2F_1\Lb\,\frac{1}{4}-\frac{1}{2}i\nu\,,\frac{3}{4}-\frac{1}{2}i\nu\,,1-i\nu\,,\frac{\Lb\,q^{\ast}\Rb\,^2}{4\Lb\,p^{\ast}\Rb\,^2}\Rb\,\notag\\
 \label{TI}
\eea

Hence, using \eq{TI} and the expression for $b_{n=0,\nu}$ in
\eq{E:c4}, the RHS of \eq{ETIL} can be written in the explicit form
as

\bea \label{FT1}\,E\Lb
\vec{p},\vec{q};\nu\Rb\,\,&=&-\frac{2^{2+2i\nu}\Lb\,p^2\Rb\,^{i\nu-\frac{1}{2}}\pi^2}{\nu}\frac{\Gamma^2\Lb\,\frac{1}{2}-i\nu\Rb\,}{\Gamma^2\Lb\,\frac{1}{2}+i\nu\Rb\,}\frac{\Gamma\Lb\,i\nu\,\Rb\,}{\Gamma\Lb\,-i\nu\Rb\,}\\
&\times\,&_2F_1\Lb\,\frac{1}{4}-\frac{1}{2}i\nu\,,\frac{3}{4}-\frac{1}{2}i\nu\,,1-i\nu\,,\frac{q^2}{4p^2}\Rb\,_2F_1\Lb\,\frac{1}{4}-\frac{1}{2}i\nu\,,\frac{3}{4}-\frac{1}{2}i\nu\,,1-i\nu\,,\frac{\Lb\,q^{\ast}\Rb\,^2}{4\Lb\,p^{\ast}\Rb\,^2}\Rb\,\notag\\
+&\,&\frac{2^{2-6i\nu}\Lb\,p^2\Rb\,^{i\nu-\frac{1}{2}}\pi^2}{\nu}\Lb\,\frac{q^2}{p^2}\Rb^{2i\nu}\,\,\,\frac{\Gamma^2\Lb\,1-i\nu\Rb\,\Gamma\Lb\,i\nu\Rb\,}{\Gamma^2\Lb\,1+i\nu\Rb\,\Gamma\Lb\,-i\nu\Rb\,}\,\,\,\,\notag\\
&\times\,&_2F_1\Lb\,\frac{1}{4}+\frac{1}{2}i\nu\,,\frac{3}{4}+\frac{1}{2}i\nu\,,1+i\nu\,,\frac{q^2}{4p^2}\Rb\,_2F_1\Lb\,\frac{1}{4}+\frac{1}{2}i\nu\,,\frac{3}{4}+\frac{1}{2}i\nu\,,1+i\nu\,,\frac{\Lb\,q^{\ast}\Rb\,^2}{4\Lb\,p^{\ast}\Rb\,^2}\Rb\,\notag
\eea

where $E\Lb \vec{p},\vec{q};n=0,\gamma\Rb$ is written as $E\Lb
\vec{p},\vec{q};\nu\Rb$.  For single pomeron exchange, (see
\fig{BFKL}), the conformal spin $\nu$ has opposite signs at the two
vertices at the ends of the pomeron. In \sec{sec:BFKL} when
\fig{BFKL} is calculated, it is assumed that $\vec{q}=0$ to simplify
the calculation. Hence, from \eq{FT1}, the product of two vertices,
(for $\vec{q}=0$) takes the form

\begin{align}
E\Lb \vec{p},\vec{q}=0;\,\nu\,\Rb\,E\Lb
\vec{p},\vec{q}=0;\,-\nu\,\Rb\,=&\frac{1}{\nu^2}\frac{16\pi^4}{p^2}\label{E:Cvert}
\end{align}

\subsection{The triple pomeron vertex}
\label{sec:3p}

\FIGURE[h]{\begin{minipage}{70mm}
\centerline{\epsfig{file=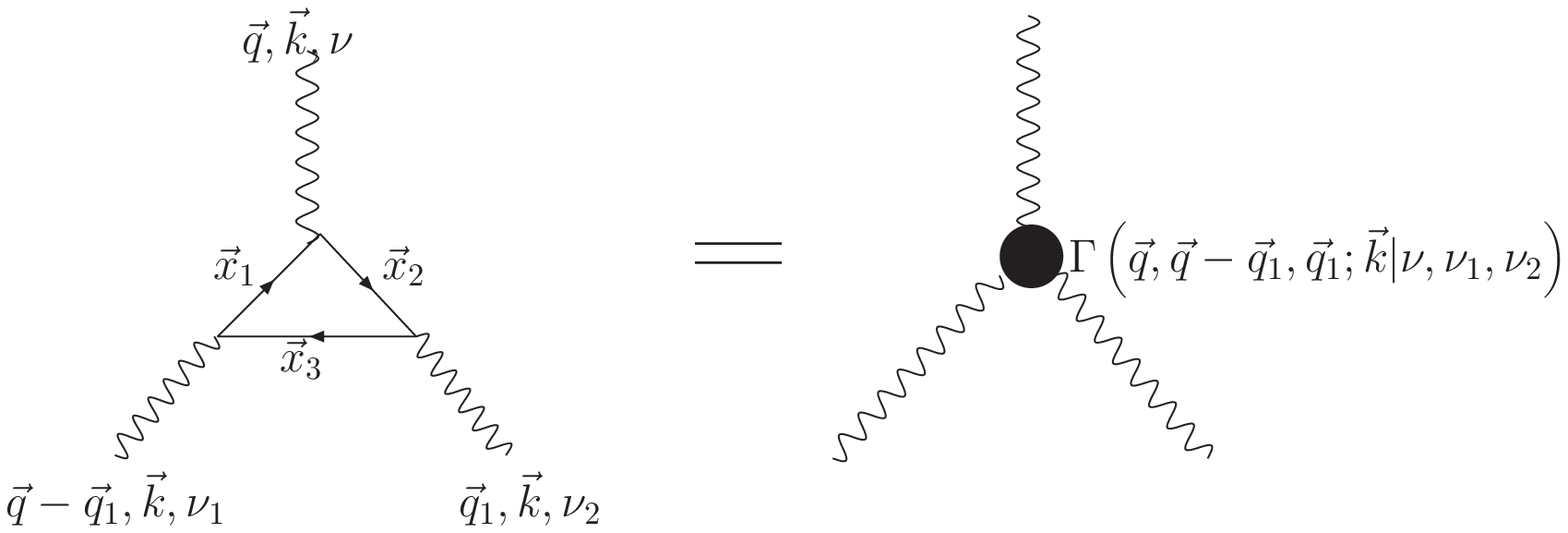,width=78mm,height=36mm}}
\end{minipage}
\caption{ The triple Pomeron vertex. } \label{vpomqk} }

In this subsection the triple pomeron vertex, illustrated in
\fig{vpomqk} is calculated explicitly in the momentum
representation. It is defined in  Refs. \cite{16,13,12}
as an integral over the centre of mass position vectors $\Lb\,x_{01},x_{02},x_{03}\Rb\,$, and the conformal dimensions $(\gamma,\gamma_1,\gamma_1)$ as\\
\\
\\

\bea
G_{3P}\Lb\vec{q},\vec{k},n=0,\gamma,\gamma_1,\gamma_2\Rb\,\,&=&\,\,G_{3P}\Lb\vec{q},\vec{k},\nu,\nu_1,\nu_2\Rb\,\,\label{E:trip}\\
&=&\,\,\int{}\frac{d^2x_{10}d^2x_{20}d^2x_{30}}{x_{12}x_{23}x_{31}}E^{n,\nu}_{q}\Lb\,x_{10},x_{20}\Rb\,E^{n,\nu_1}_{k}\Lb\,x_{20},x_{30}\Rb\,E^{n,\nu_2}_{q-k}\Lb\,x_{30},x_{10}\Rb\nonumber
\eea

To calculate the triple pomeron vertex explicitly, the mixed
representation of Lipatov in Ref.\cite{11} is used for the vertex
eigenfunctions $E^{n,\nu}_{q}$ (see \eq{E:cc3}). Note that to
simplify the calculation of the first enhanced diagram of
\fig{BFKLenh}, it is assumed that $\vec{q}=0$ for the momentum
transferred along the pomeron, above and below the pomeron loop.
Hence, the triple pomeron vertex shown in \fig{vpomqk} is calculated
for $\vec{q}=0$. In Ref.\cite{12} this mixed representation was used
in the definition of \eq{E:trip} to give the expression

\begin{equation}
G_{3P}(\vec{q}=0,\vec{k},\nu{},\nu_{1},\nu_{2})=\frac{1}{2\pi{}}\int\frac{d^{2}x_{01}}{x_{01}^{2}}x_{01}^{-2i\nu-1}e^{\frac{i\vec{k}\cdot\,\vec{x}_{01}}{2}}\int{}d^{2}x_{2}\frac{x_{01}^{2}}{x_{12}^{2}x_{02}^{2}}
x_{02}^{2i\nu_1+1}E_{k}^{n,\nu_1}(x_{02})x_{12}^{2i\nu_2+1}E_{-k}^{n,\nu_2}(x_{12})\label{E:3.1.8}
\end{equation}

It is also assumed that $\nu_1=\nu_2=0$ in evaluating the expression
of \eq{E:3.1.8} for the triple pomeron vertex, for the following
reason. When evaluating the integrals over $\nu_1$ and $\nu_2$, for
the expression of \fig{BFKLenh}, (see \eq{E:3.1.7}), one expands the
BFKL functions $\omega\Lb\nu_1\Rb\,$ and $\omega\Lb\,\nu_2\Rb\,$
around the saddle point $\nu_1=\nu_2=0$ (see \eq{E:A19}), which
gives the largest contribution to the integration. In the appendix,
the integral of \eq{E:3.1.8} is evaluated to give the triple pomeron
vertex as an explicit expression in the momentum representation in
\eq{E:A12} as

\begin{align}
&G_{3P}(\vec{q}=0,\vec{k},\nu,\nu_{1}\rightarrow\,0,\nu_{2}\rightarrow\,0,)\,\,\,=\,\,\,k^{i\nu-\frac{1}{2}}\frac{2^{-2i\nu}}{4\nu_{1}\nu_2\pi}\frac{\Gamma^3\Lb\,\frac{1}{2}-i\nu\Rb\,\Gamma^2\Lb\,i\nu\Rb\,}{\Gamma\Lb\,\frac{1}{2}+i\nu\Rb\,}\label{E:3.1.11}
\end{align}

\subsection{The single pomeron amplitude}
\label{sec:BFKL}

\FIGURE[h]{ \epsfig{file= 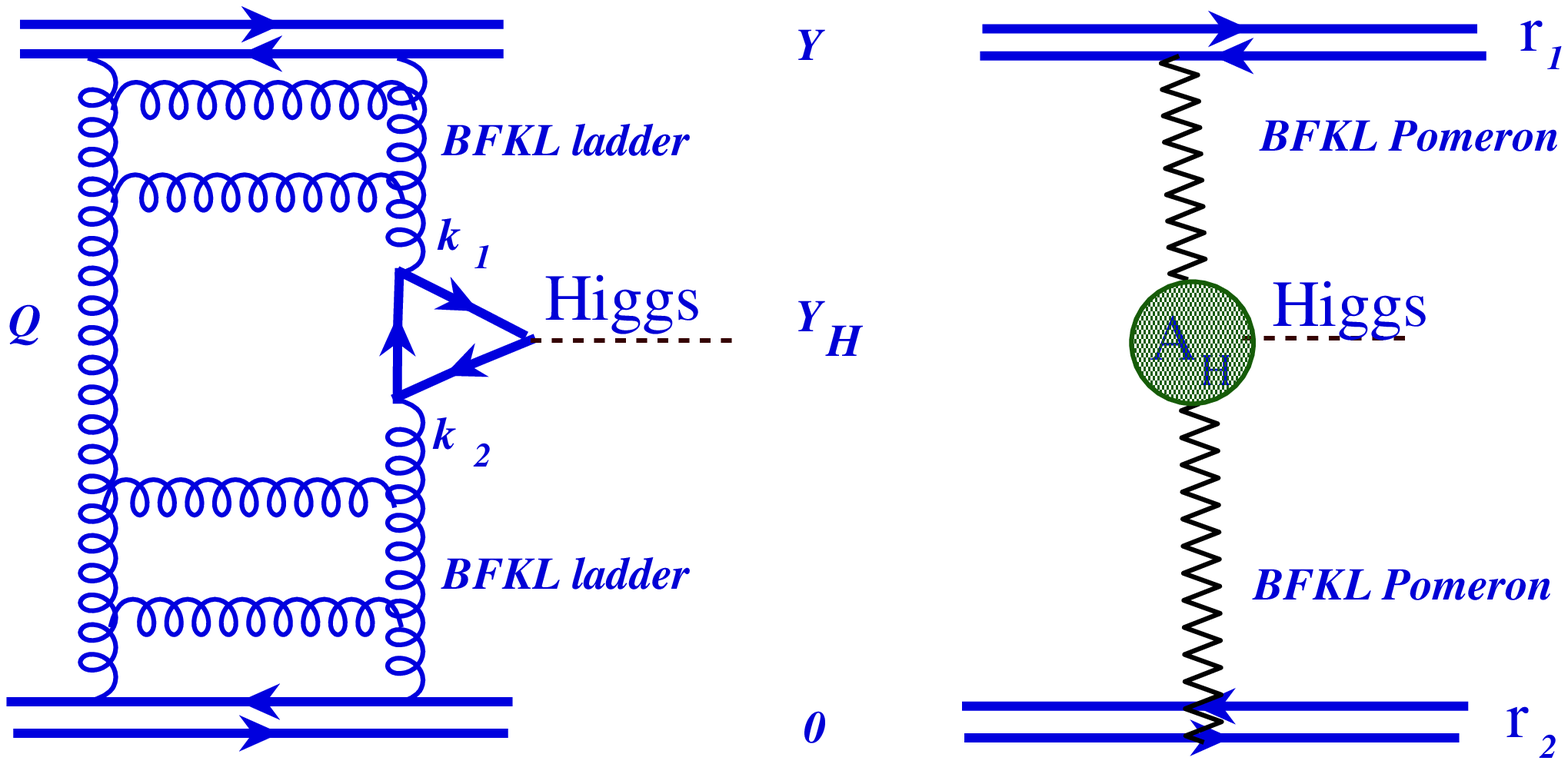,width=120mm,height=60mm}
\caption{Central diffractive production in colour dipole scattering
due to single pomeron exchange. } \label{BFKL} }

In this subsection the single pomeron amplitude with Higgs
production shown in \fig{BFKL} is calculated. The two dipoles are
separated by a rapidity gap $Y$, and they have transverse sizes
$r_{1}=x_{12}$ and $r_{2}=x_{12}'$. The momenta conjugate to the
dipole sizes are $\vec{p}_{1}$ and $\vec{p}_{2}$, and $\vec{q}$ is
the momentum transferred along the pomeron. For simplicity it is
assumed that $\vec{q}=0$.
The single pomeron
amplitude  with Higgs production, in the QCD dipole approach is denoted $M_{Higgs}\Lb\,n=1\,,\,Y\Rb\,$, where $Y$ is the rapidity gap between the two incoming protons, and $n=1$ denotes the single pomeron exchanged between the two protons. In this notation, the single pomeron amplitude, with Higgs production of \fig{BFKL}, has the expression
\cite{11,16,15,14}

\beq
M_{Higgs}\Lb\,n=1\,,\,Y\Rb\,\,=\,\,P^{BFKL}(\vec{p}_{1},\vec{p}_{2},Y,\vec{q}=0)\,A_{H}\Lb\,\delta\,Y_H\Rb\,\label{MBFKL}
\eeq

where $P^{BFKL}(\vec{p}_{1},\vec{p}_{2},Y,\vec{q}=0)\,$ is the single pomeron amplitude given by the expression
\begin{equation}
P^{BFKL}(\vec{p}_{1},\vec{p}_{2},Y,\vec{q}=0)=\frac{\alpha_{s}^{2}}{4}\int{}\frac{d\nu{}}{2\pi{}i}\mathcal{D}(\nu{})e^{\omega(\nu{})}E\Lb\,\vec{p}_1,\vec{q}=0,\nu{}\Rb\,E\Lb\,\vec{p}_2,\vec{q}=0,-\nu{}\Rb\,\label{E:3.1.2}
\end{equation}

 and where $A_{H}\Lb\,\delta\,Y_H\Rb\,$ denotes the subprocess contribution
which produces the Higgs,
such as the quark triangle subprocess of \fig{tr}. The typical rapidity window, which the Higgs
 Boson occupies is $\delta\,Y_H\,=\,\ln\Lb\,\frac{M_H^2}{4m^2}\Rb\,$, where $m$ is the mass of the proton.
 The simplest subprocess with the largest contribution for Higgs production in the standard model, is the quark triangle shown in \fig{tr}. After the
subprocess amplitude of \fig{tr} is contracted with the gluon propagators, the expression for the contribution of the quark triangle shown in \fig{tr} is given by \cite{19,31}

\FIGURE[h]{ \centerline{\epsfig{file=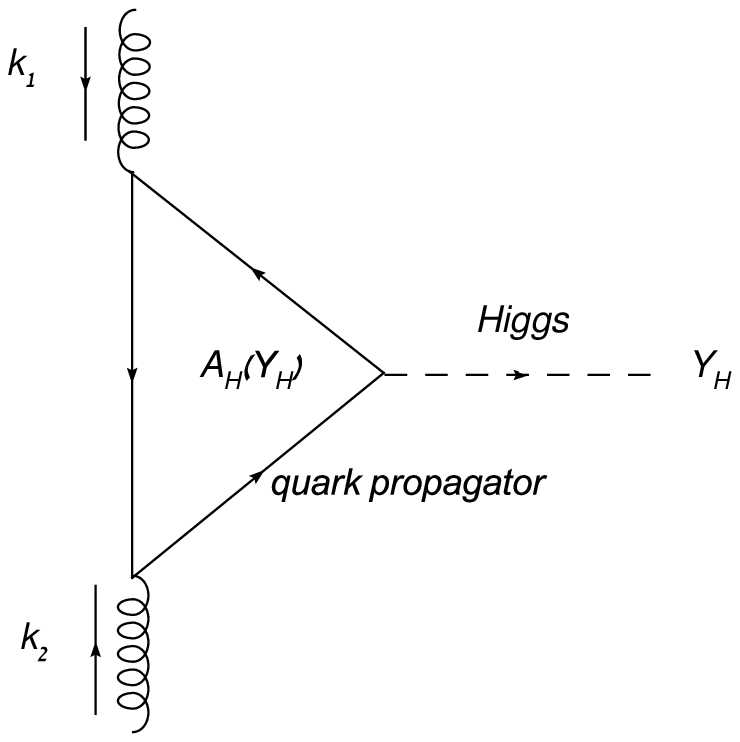,width=70mm,height=70mm}}
\caption{Quark triangle subprocess for Higgs production.} \label{tr} }

\beq
A_H\Lb\,\delta\,Y_H\Rb\,\,=\,\,A\Lb\,M_H^2\Rb\,\,\Lb\,\vec{k}_{1}\cdot\,\vec{k}_{2}\,\Rb\,\,\label{triangle}
\eeq

where the factor $A\Lb\,M_H^2\Rb\,$ has the value \cite{Higgs,Rizz,daw,22}

\begin{align}
A\Lb\,M_H^2\Rb\,=\,\frac{2}{3}\left(\frac{-\alpha{}_{s}\left(M_{H}^{2}\right)\left(\sqrt{2}G_{F}\right)^{\frac{1}{2}}}{\pi{}}\right)\label{E:A}
\end{align}

where $G_F$ is the Fermi coupling. $D\left(\nu\right)$ appearing in the single pomeron amplitude of \eq{E:3.1.2} is given by
\begin{equation}
\mathcal{D}(\nu{})=\frac{\nu{}^{2}}{\Lb\,\nu{}^{2}+\frac{1}{4}\Rb\,^{2}}\label{E:3.1.3}
\end{equation}

$\omega(n=0,\nu{})$ is the solution to the BFKL equation defined in
\eq{OM}, where in the high energy limit one takes $n=0$. From now on
 the notation $\omega(n=0,\nu{})=\omega\Lb\,\nu\Rb$ is used. Assuming that the
conjugate momenta $\vec{p}_1$ and $\vec{p}_2$ of the two scattering
dipoles in \fig{BFKL} are equal in magnitude, \eq{E:Cvert} can be
used for the product of the two pomeron vertices. Hence,
\eq{E:3.1.3} can be written as

\begin{equation}
P^{BFKL}(\vec{p}_{1}=\vec{p}_{2}=\vec{p},Y,\vec{q}=0)=\frac{4\alpha_{s}^{2}\pi^4}{p^2}\int{}\frac{d\nu{}}{2\pi{}i}\frac{1}{\Lb\,\frac{1}{4}+\nu^2\Rb^2}\,e^{\omega(\nu{})Y}\label{E:3.2.3}
\end{equation}

The integration over $\nu$ can be evaluated at the saddle point
$\nu=0$ of $\omega\Lb\nu\Rb$. In this way, the RHS of \eq{E:3.2.3}
becomes

\begin{align}
P^{BFKL}(\vec{p}_{1}=\vec{p}_{2}=\vec{p},Y,\vec{q}=0)&=\frac{32\as^2\pi^3}{p^2}\Lb\,\frac{2\pi{}}{(\omega{}"(\nu{}=0)Y)}\Rb^{\frac{1}{2}}\,e^{\omega{}(\nu{}=0)Y}\label{E:3.2.3aaa}
\end{align}

Hence, the final expression of \eq{MBFKL} for the process of \fig{BFKL} reads

\begin{align}
M_{Higgs}\Lb\,n=1\,,\,Y\Rb\,&=\frac{32\as^2\pi^3}{p^2}\Lb\,\frac{2\pi{}}{(\omega{}"(\nu{}=0)Y)}\Rb^{\frac{1}{2}}\,
e^{\omega{}(\nu{}=0)Y}\,A_H\Lb\,\delta\,Y_H\Rb\,\label{MBFKL1}
\end{align}

This is the expression for the single pomeron amplitude, including Higgs production, of
\fig{BFKL}. However, \eq{MBFKL1} is  written in the approximation that $s  = (p_1 + p_2)^2 \,\gg\,M^2_H$ . Since we expect the Higgs mass to be large, we take into account the main correction due to this mass, namely, we make the following replacement.
\beq \label{MHCOR}
e^{\omega{}(\nu{}=0)Y}\,\equiv\,\,\left( \frac{s}{m^2} \right)^{\omega(\nu=0)\,Y}\,\,\longrightarrow\,\,
\left( \frac{1}{x_1\,x_2} \right)^{\omega(\nu=0)\,Y}
\left( \frac{4\,s}{M^2_H} \right)^{\omega(\nu=0)\,Y}
\,\,\equiv\,\,e^{\omega{}(\nu{}=0)\,\Lb\,Y -  \ln(M^2_H/4m^2)\Rb\,}
\eeq
where $m$ is the mass of proton., $x_1$ and $x_2$ are equal to $k^2_1/s_1$ and $k^2_2/s_2$ (see \fig{BFKL}) with $s_1 = (p_1 + k_1)^2 $ and $s_2 = (p_2 +k_2)^2$). Using the well known kinematic relation $s_1\,s_2 \,=\,M^2_H\,s$ and
since   $k^2_1 \,=\,k^2_2\,=\,M^2_H/2$ (see Ref. \cite{durham}
for example). Finally, \eq{MBFKL1} looks as follows
\begin{align}
 M_{Higgs}\Lb\,n=1\,,\,Y\Rb\,&=\frac{32\as^2\pi^3}{p^2}\Lb\,\frac{2\pi{}}{(\omega{}"(\nu{}=0)Y)}\Rb^{\frac{1}{2}}\,
e^{\omega{}(\nu{}=0)\,(Y\,-\,\ln(M^2_H/4m^2))}\,A_H\Lb\,\delta\,Y_H\Rb\,\label{MBFKL2}
\end{align}

As one can see in \eq{MBFKL2} the single Pomeron exchange does not depend on the value of Higgs boson rapidity ($Y_H$)
but depends on $\delta Y_H = \ln(M^2_H/4m^2)$ which characterizes the window in rapidity occupied by the heavy Higgs boson.
\subsection{The first enhanced amplitude}
\label{sec:BFKLenh}

In this subsection the amplitude for the first enhanced amplitude,
with Higgs production shown in \fig{BFKLenh}, is calculated. The
pomeron loop is between the two rapidity values $Y_{1}$ and $Y_{2}$.
Hence, one needs to integrate over these two rapidity values. There
is also an integral to evaluate, over the unknown momentum $\vec{k}$
in the pomeron loop. The enhanced diagram
with Higgs production, in the QCD dipole approach is denoted $M_{Higgs}\Lb\,n=2\,,\,Y\Rb\,$, where $n=2$  denotes the splitting of the exchanged pomeron, into two branches forming the loop in \fig{BFKLenh}. The amplitude of \fig{BFKLenh}, is the first hard rescattering correction, to the single pomeron amplitude of \fig{BFKL}. In this notation, first enhanced amplitude, with Higgs
production of \fig{BFKLenh} is given by

\beq
M_{Higgs}\Lb\,n=2\,,\,Y\Rb\,\,=\,\,2\,P^{BFKL}_{enhanced}(\vec{p}_{1},\vec{p}_{2},Y,\vec{q})\,A_H\Lb\,\delta\,Y_H\Rb\,\label{MBFKLenh}
\eeq

where $P^{BFKL}_{enhanced}(\vec{p}_{1},\vec{p}_{2},Y,\vec{q})$ is the BFKL pomeron amplitude for the first enhanced one - loop diagram, which has the expression given below in \eq{E:3.1.7}. The factor of $2$ in \eq{MBFKLenh}, comes from adding the two identical contributions of \fig{BFKLenh}, due to the two ways the Higgs is emitted from the two branches of the pomeron loop. In order to obtain the complete contribution of \fig{BFKLenh}, both possibilities for Higgs production from the two branches of the loop must be considered separately, and added. $P^{BFKL}_{enhanced}(\vec{p}_{1},\vec{p}_{2},Y,\vec{q})$ is given by the expression (see Refs. \cite{16,15})

\begin{align}
P^{BFKL}_{enhanced}(\vec{p}_{1},\vec{p}_{2},Y,\vec{q})=&\,\,\,B\int\!d\nu\,d\nu'd\nu_1d\nu_2d^2k{}\int^{Y}_{Y_H
+ \frac{1}{2} \delta Y_H\,}dY_{1}\int^{Y_H  - \frac{1}{2} \delta Y_H}_{0}dY_{2}\notag\\
&\times\,E\Lb\,\vec{p}_1,\vec{q},\nu{}\Rb\,\mathcal{D}\Lb\,\nu\Rb\,e^{\omega{}(\nu{})(Y-Y_{1})}G_{3P}(\nu,\nu_{1},\nu_{2},\vec{q}=0,\vec{k})\notag\\
&\times\,\mathcal{D}\Lb\,\nu_1\Rb\,\mathcal{D}\Lb\,\nu_2\Rb\,e^{(\omega(\nu_{1})+\omega(\nu_{2}))(Y_{1}-Y_{2} ) \, - \,\omega(\nu_{2})\,\ln(M^2_H/4m^2)} \,G_{3P}(-\nu{}',\nu_{1},\nu_{2},\vec{q}=0,\vec{k})\notag\\
&\times\,e^{\omega{}(\nu{}')Y_{2}}E\Lb\,\vec{p}_2,\vec{q},-\nu{}'\Rb\,\label{E:3.1.7}\\
\mbox{where}\,\,\,\,\, B=&\,\,\,-
\frac{\as^4\pi^4}{8}\Lb\,\frac{\as\,N_c}{2\pi^2}\Rb^2\label{B}
\end{align}

where $Y_H$ is the rapidity of the Higgs boson. $Y_H$ here is considered equal to zero in the c.m. frame, restricting ourselves to the production of the Higgs boson at rest in the c.m. frame
 since it is the most likely experimental kinematic,  and $\delta Y_H = \ln\left(M^2_H/4 m^2\right)$  characterizes the rapidity window, occupied by the Higgs boson.
 \FIGURE[ht]{ \epsfig{file=
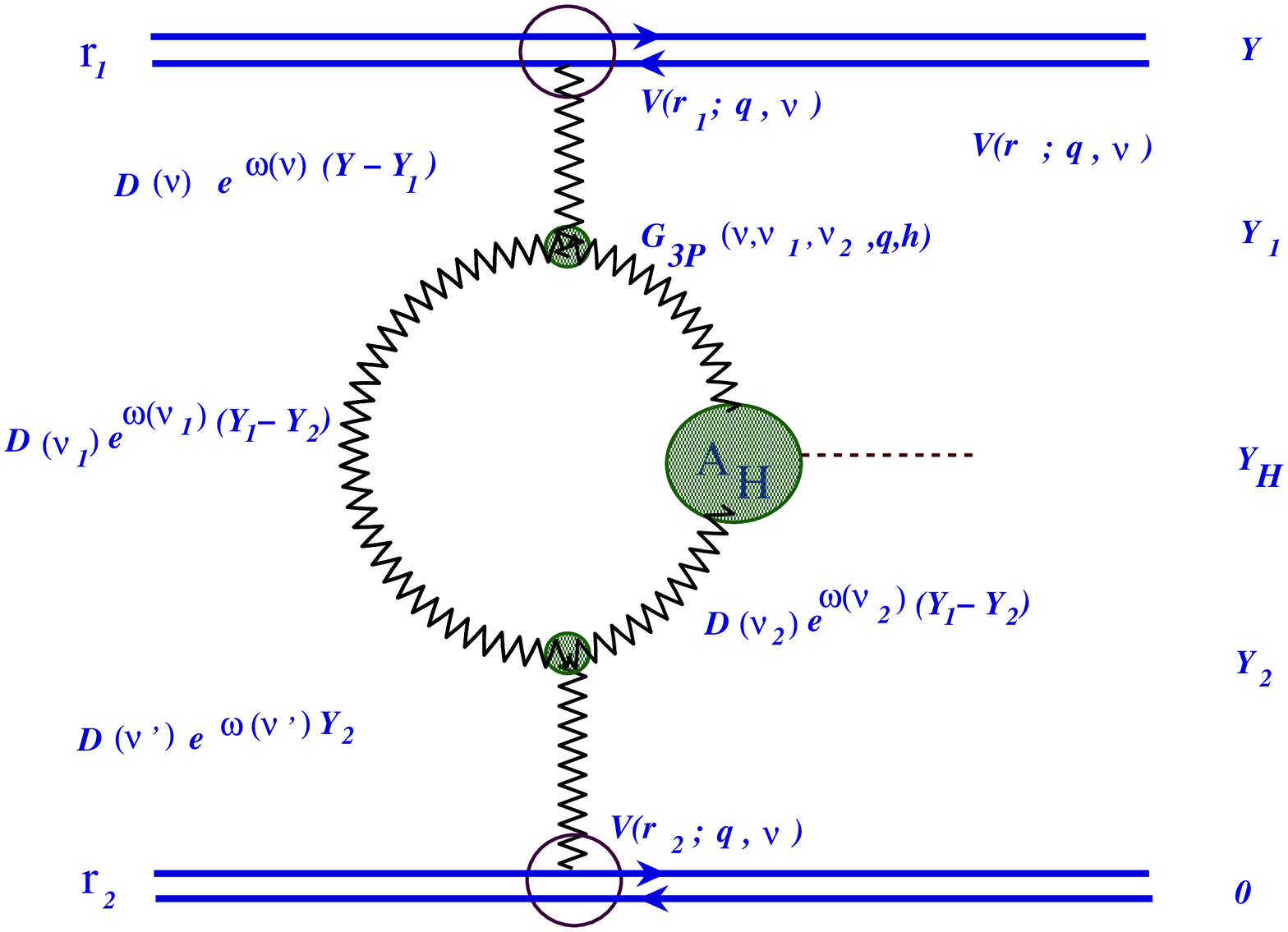,width=120mm,height=90mm} \caption{Central diffractive
production in colour dipole scattering due to pomeron exchange with
a hard rescattering correction. } \label{BFKLenh} }

Using the same assumptions as \sec{sec:3p}, the integral over $\vec{k}$
in \eq{E:3.1.7} is evaluated in the appendix with the result given
in \eq{E:A13} as

\bea
&\,\,&\int\!d^{2}kG_{3P}(\nu,\nu_{1}\rightarrow\,0,\nu_{2}\rightarrow\,0,\vec{q}=0,\vec{k})G_{3P}(-\nu{}',\nu_{1}\rightarrow\,0,\nu_{2}\rightarrow\,0,\vec{q}=0,\vec{k})\notag\\
&\xrightarrow{\nu_1,\nu_2\rightarrow\,0}&\frac{2^{2(i\nu{}'-i\nu{})}}{8}\frac{\delta{}(\nu{}-\nu{}')}{4\nu_{1}^{2}4\nu_{2}^{2}}\frac{\Gamma^3\Lb\,\frac{1}{2}+i\nu{}'\Rb\,\Gamma^3\Lb\,\frac{1}{2}-i\nu{}\Rb\,\Gamma^2\Lb\,-i\nu'\Rb\,\Gamma^2\Lb\,i\nu\Rb\,}{\Gamma\Lb\,\frac{1}{2}-i\nu{}'\Rb\,\Gamma\Lb\,\frac{1}{2}+i\nu{}\Rb\,}\label{E:3.2.6}
\eea

Note that the $\nu_1^2$ and $\nu_2^2$ in the denominator of
\eq{E:3.2.6} cancel with $\mathcal{D}\Lb\,\nu_1\Rb\,$ and
$\mathcal{D}\Lb\,\nu_2\Rb\,$ in \eq{E:3.1.7} (see \eq{E:3.1.3}).
When inserting \eq{E:3.2.6} into the right hand side of
\eq{E:3.1.7}, the delta function allows the integration over
$\nu{}'$ to be evaluated to give the result

\begin{align}
P^{BFKL
}_{enhanced}(\vec{p}_{1},\vec{p}_{2},Y,\vec{q}=0)=&\,\,\,\frac{B\pi^2}{8}\int\!d\nu\,d\nu_{1}d\nu_{2}\int^{Y}_{Y_H  + \h \delta Y_H}dY_{1}\int^{Y_{H} - \h \delta Y_H}_{0}dY_{2}\notag\\
&\times\,E\Lb\,\vec{p}_1,\vec{q},\nu{}\Rb\,\mathcal{D}^{2}(\nu)e^{\omega(\nu)(Y-Y_{1}+Y_{2})}\frac{\Gamma{}^{2}\Lb\,\frac{1}{2}-i\nu{}\Rb\,\Gamma{}^{2}\Lb\,\frac{1}{2}+i\nu{}\Rb\,}{4\nu_{1}^{2}4\nu_{2}^{2}\nu{}^{2}\sin{}^{2}\left(i\nu{}\pi{}\right)}\notag\\
&\times\mathcal{D}(\nu_{1})\mathcal{D}(\nu_{2})e^{(\omega(\nu_{1})+\omega(\nu_{2}))(Y_{1}-Y_{2})\,\,-\,\,\omega(\nu_2)\,\ln(M^2_H/4m^2)}\,E\Lb\,\vec{p}_2,\vec{q},-\nu{}\Rb\,\label{E:3.2.7}
\end{align}

Now the integrals over $\nu$,$\nu_{1}$,$\nu_{2}$ and the two
rapidity values $Y_{1}$ and $Y_{2}$ need to be evaluated. $Y_1$ and
$Y_2$ are the upper and lower rapidity values for the pomeron loop
in \fig{BFKLenh}. The details of the integrations are given in the
appendix in \eq{E:A14}-\eq{E:A20}, and the final expression is given
in \eq{E:A20} as

\bea
&\,&P^{BFKL}_{enhanced}(\vec{p}_{1},\vec{p}_{2},Y,\vec{q}=0)=\notag\\
&\,&\frac{32\,B\pi^8\,}{\Lb\,2\bas\Rb^5p^2}\frac{\delta\,Y_H}{\omega"\Lb\,\nu\,=0\Rb\,}\Lb\,\frac{\Lb\,2\omega\Lb\,\nu\,=0\,\Rb\,\Rb^4\,}{Y}-4\frac{\Lb\,2\omega"\Lb\,\nu\,=0\,\Rb\,\Rb^3}{Y^2}\Rb\,\,e^{2\omega\Lb\,\nu\,=0\,\Rb\,\Lb\,Y-\h\,\ln\Lb\,\frac{M_H^2}{4m^2}\Rb\,\Rb\,}\,
\label{E:3.2.13}
\eea

where the constant $B$ is given in \eq{B}. Therefore, the full expression for the diagram of \fig{BFKLenh} given by \eq{MBFKLenh} takes the form

\bea
&\,&M_{Higgs}\Lb\,n=2\,,\,Y\Rb\,\,=\,\label{MBFKLenh1}\\
&\,&\frac{64\,B\pi^8\,}{\Lb\,2\bas\Rb^5p^2}\frac{\delta\,Y_H}{\omega"\Lb\,\nu\,=0\,\Rb\,}\Lb\,\frac{\Lb\,2\omega\Lb\,\nu\,=0\,\Rb\,\Rb^4\,}{Y}-4\frac{\Lb\,2\omega"\Lb\,\nu\,=0\,\Rb\,\Rb^3}{Y^2}\Rb\,\,e^{2\omega\Lb\,\nu\,=0\,\Rb\,\Lb\,Y-\h\,\ln\Lb\,\frac{M_H^2}{4m^2}\Rb\,\Rb\,}\,A_{H}\Lb\,\delta\,Y_H\Rb\,\notag
\eea

\section{The survival probability in diffractive Higgs production in colour dipole scattering due to pomeron exchange}
\label{sec:sp}
\subsection{The definition of survival probability}
\label{sec:spdef}

In this section the survival probability of large rapidity gaps, in diffractive Higgs production is
calculated, using the ratio of \fig{BFKL} and \fig{BFKLenh} (see
and \eq{MBFKL1} and \eq{MBFKLenh1}). To guarantee that there will still be a
 large rapidity gap (LRG) between the protons after scattering,
 all hard rescattering corrections, that could give terms filling up the LRG,
  must be taken into account. The survival probability, is the probability to
   just have the exclusive Higgs production shown in \fig{BFKL}, and not
    to have any higher order hard rescattering corrections, such as the first
  enhanced diagram of \fig{BFKLenh}. In other words , the survival probability is the ratio of the  calculated cross section for the Higgs boson production to the one form the one pomeron exchange.
Hence, the survival probability of the LRG,
  is calculated by subtracting the sum over all hard rescattering amplitudes
   from the single pomeron amplitude of \fig{BFKL}, and dividing the result
    by the single pomeron amplitude of \fig{BFKL} itself, to obtain
    the correctly normalised survival probability. Therefore, the survival
     probability is defined as

\beq
<\,\vert\,S^2\vert\,>\,=\,\frac{M_{Higgs}\Lb\,n=1\,,\,Y\Rb\,-\sum^{\infty}_{n=2}\,\Lb\,-1\Rb^n\,M_{Higgs}\Lb\,n\,,\,Y\Rb\,}{M_{Higgs}\Lb\,n=1\,,\,Y\Rb\,}   \label{defsp}
\eeq

where $M_{Higgs}\Lb\,n\,,\,Y\Rb\,$ is the $n^{th}$ order hard rescattering correction. For example, in the case of $n=2$, the first hard rescattering correction $M_{Higgs}\Lb\,n=2\,,\,Y\Rb\,$ is the contribution of the first enhanced diagram of \fig{BFKLenh}, which has $2$ pomeron branches, forming the pomeron loop. In general, $M_{Higgs}\Lb\,n\,,\,Y\Rb\,$ is the contribution given by the diagram which has $n$ pomeron branches. In calculating the survival probability, if only the first enhanced diagram is taken into account, and corrections of the order $n=3$ and higher are ignored, then the formula of \eq{defsp} reduces to

 \beq
 <\,\vert\,S^2\vert\,>\,=\,\frac{M_{Higgs}\Lb\,n=1\,,\,Y\Rb\,-\,M_{Higgs}\Lb\,n=2\,,\,Y\Rb\,}{M_{Higgs}\Lb\,n=1\,,\,Y\Rb\,}\,=\,1-\frac{M_{Higgs}\Lb\,n=2\,,\,Y\Rb\,}{M_{Higgs}\Lb\,n=1\,,\,Y\Rb\,}   \label{defsp1}
 \eeq

 The ratio $\frac{M_{Higgs}\Lb\,n=2\,,\,Y\Rb\,}{M_{Higgs}\Lb\,n=1\,,\,Y\Rb\,}$ is calculated in the next subsection, in the symmetric QCD dipole approach (see \eq{E:3.2.16} in \sec{sec:sp2}). It turns out
that this ratio is not small and, therefore, all enhanced diagrams
need to be taken into account. Using the toy model suggested by
Mueller in Ref.\cite{toy}, all enhanced diagrams are taken into
account in the Mueller - Patel- Salam - Iancu (MPSI) approach (see
Refs.\cite{8,9,10}). The formula for the scattering amplitude in
this
model was suggested by Kovchegov in Ref. \cite{20}.\\

\numberwithin{equation}{subsection}

\subsection{The QCD dipole approach}
\label{sec:sp2}

The survival probability of large rapidity gaps, in diffractive Higgs production in the QCD dipole approach, is the probability for the exclusive Higgs production of \fig{BFKL}, with a large rapidity gap between the Higgs signal and the two emerging dipoles. To calculate the survival probability, all hard rescattering corrections which could fill up the large rapidity gaps must be subtracted from the single BFKL Higgs amplitude $M_{Higgs}\Lb\,n=1\,,\,Y\Rb\,$, and the result must be divided by $M_{Higgs}\Lb\,n=1\,,\,Y\Rb\,$. If only the first enhanced rescattering correction $M_{Higgs}\Lb\,n=2\,,\,Y\Rb\,$ is
taken into account, then the survival probability in the symmetric QCD dipole approach is estimated as

\bea
<|S^2|>\,=\,\frac{M_{Higgs}\Lb\,n=1\,,\,Y\Rb\,-M_{Higgs}\Lb\,n=2\,,\,Y\Rb\,}{M_{Higgs}\Lb\,n=1\,,\,Y\Rb\,}\,\,=\,1\,-\,\frac{M_{Higgs}\Lb\,n=2\,,\,Y\Rb\,}{M_{Higgs}\Lb\,n=1\,,\,Y\Rb\,}\label{sp2}\eea

where the amplitudes $M\Lb\,n=1\,,\,Y\,\Rb\,$ and
$M\Lb\,n=2\,,\,Y\Rb$ have been calculated in \eq{MBFKL1} and
\eq{MBFKLenh1}, respectively. Using the results of \eq{MBFKL1}
and \eq{MBFKLenh1}, then the ration $\frac{M\Lb\,n=2\,,\,Y\,\Rb\,}{M\Lb\,n=1\,,\,Y\,\Rb\,}$ appearing in \eq{sp2} is found to have the expression

\bea
&\,&\frac{M_{Higgs}\Lb\,n=2\,,\,Y\Rb\,}{M_{Higgs}\Lb\,n=1\,,\,Y\Rb\,}=\label{E:sp}\\
&\,&\frac{2\,B\pi^5\,}{\as^2\,\Lb\,2\bas\Rb^5}\frac{\delta\,Y_H}{\omega\,"\Lb\,\nu\,=0\,\Rb\,}\Lb\,\frac{\Lb\,2\omega\Lb\,\nu=\,0\Rb\,\Rb^4}{Y}-\,4\,\frac{\Lb\,2\omega\Lb\,\nu\,=\,0\Rb\,\Rb^3}{Y^2}\Rb\,\Lb\,\frac{\omega"\Lb\,\nu\,=0\,\Rb\,Y}{2\pi}\Rb^{\frac{1}{2}}e^{\omega\Lb\,\nu\,=0\,\Rb\,Y}
\notag
\eea

where the constant $B$ is given in \eq{B}. Here a typical value for
$\alpha{}_{s}$, which depends on the mass of the $Z$ particle, is
used. It is expected that the Higgs will be produced with a mass of
approximately $100$ $GeV$, which would give a value for the strong
coupling constant $\alpha{}_{s}\sim{}0.12{}$. This corresponds to a
$Z$ particle mass \cite{25}, of $M_{Z}=90.8\pm{}0.6
$ $Gev$.\\
The following values are to be found in Ref. \cite{25} and
Ref. \cite{26}, for the strong coupling and the BFKL function.

\begin{align}
\alpha_{s}=0.12& &\omega{}(\nu{}=0)=\bar{\alpha}_{s}4\ln2&
&\frac{1}{2}\omega{}"(\nu{}=0)=14\bas\,\zeta(3)&
&\zeta(3)\cong{}1.202\label{E:3.2.15}
\end{align}

Assuming that the rapidity gap $Y$ at the LHC is 19, and using the
numerical values given in \eq{E:3.2.15}, the right hand side of
\eq{E:sp} then yields following.

\begin{equation}
\frac{M\Lb\,n=2\,,\,Y\,\Rb\,}{M\Lb\,n=1\,,\,Y\Rb\,}=2.8\,\,\,e^{\omega{}(\nu{}=0)Y}\label{E:3.2.16}
\end{equation}

This value is not small and increases with energy. Therefore, it
shows that all enhanced diagrams have to be taken into account. In
the next section all enhanced diagrams are summed in the toy model.

\subsection{The toy model approach} \label{sec:toy}

In this subsection, the survival probability is calculated taking
into account all enhanced diagrams. The toy model proposed by
Mueller in Ref. \cite{toy}, is a model for describing pomeron
exchange in onium - onium scattering. In the toy model, the dipole
wave function of an onium is described by the generating functional
for dipoles (\cite{toy}) $Z(Y ,[u])$:
{
\beq \label{LD1}
Z\left(Y\,-\,Y_0;\,[u] \right)\,\,\equiv\,\,
\eeq
$$
\equiv\,\,\sum_{n=1}\,\int\,\,
P_n\left(Y;\,r_1, b_1,\,r_2, b_2, \dots ,r_i, b_i, \dots ,r_n, b_n
 \right) \,\,
\prod^{n}_{i=1}\,u(r_i, b_i) \,d^2\,r_i\,d^2\,b_i
$$
Here, $P_n$ are the probabilities to find dipoles with sizes $r_i$ and impact parameters $b_i$ at rapidity Y and  $u(x_{01},b)$  is
an arbitrary  function of the dipole of transverse size $x_{01}$, at impact
parameter $b$ .  In the toy model which we are going to consider here  we neglect the dependence of $u$ on the size of dipoles and their impact parameters(see Ref. \cite{toy}) . In this model
 $Z(x_{01},b,Y,u)$ degenerates to the generating function and  obeys the following evolution equation (see Ref.
\cite{toy})

\begin{equation}
\frac{dZ\Lb\,Y,u\Rb\,}{dY}=\Delta\,Z^{2}\Lb Y,u\Rb\,-\Delta\,Z\Lb Y,u\Rb\,\label{E:4.1}
\end{equation}

where $\Delta\,$ is the pomeron intercept. In \sec{sec:BFKL} and
\sec{sec:BFKLenh}, the pomeron intercept  can be  taken to be the BFKL
intercept $\Delta\,=\,\omega\Lb\,\nu\,=\,0\Rb\,$ to provide a matching with the BFKL Pomeron calculus}. The initial
condition for \eq{E:4.1} is given by

\begin{equation}
Z(Y=0,u)=u\,\label{E:4.2}
\end{equation}

The solution of the toy model \eq{E:4.1}, which satisfies the
initial condition of \eq{E:4.2} is \cite{toy,20}

\begin{equation}
Z(Y,u)=\frac{u\,}{u\,+(1 - u)\,e^{\Delta\,Y}}\label{E:4.3}
\end{equation}

\eq{E:4.3} gives the sum over all "fan" diagrams. To generalise this
result to the sum over all essential enhanced diagrams, the MPSI
approximation is used to sum over all diagrams, with pomeron loops
larger than $\frac{Y}{2}$. In Ref.\cite{20}, the forward scattering
amplitude in the MPSI approximation was written and has the form
\begin{equation}
D(Y\,,\,d)=1-\exp\Lb\,-d\frac{d{}^{2}} {d{}u\,d{}v}\Rb\,Z\left(\frac{Y}{2},u\right)Z\left(\frac{Y}{2},v\right)\vert_{u=1,v=1}\label{E:4.4}
\end{equation}

where $d$ is the dipole amplitude $(0<\,d<\,1)$ at low energy. Substituting for $Z\left(Y,u\right)$, the right hand side
of \eq{E:4.3} in equation \eq{E:4.4}, yields the following
expression for $D(Y\,,\,d)$ \cite{20}

\beq
D\Lb\,Y\,,\,d\Rb\,=\,-\,\sum_{n=1}^{\infty}\,\Lb\,-1\Rb^nD\Lb\,n\,,\,Y\,,\,d\Rb\,\,=\,-\sum^{\infty}_{n=1}n!\Lb\,-1\Rb^nd^n\,e^{\,n\Delta\,Y}\,\Lb\,1-e^{\,\Delta\,\,\frac{Y}{2}}\Rb^{2n-2}\,\label{kovchegov}\eeq

At large rapidity values, one can make the approximation
$1-e^{\,\Delta\,\frac{Y}{2}}\,\,\approx\,-e^{\,\Delta\,\,\frac{Y}{2}}$,
such that \eq{kovchegov} can be re-written as

\begin{equation}
D(Y\,,\,d)\,=\,-\sum^{\infty}_{n=1}\Lb\,-1\Rb^n\,D\Lb\,n\,,\,Y\,,\,d\Rb\,\,=\,-\sum^{\infty}_{n=1}n!(-1)^{n}d^{n}e^{n\Delta\,Y}\label{E:4.5}
\end{equation}

In \eq{E:4.5}, the $n^{th}$ term is the amplitude for $n$ - pomeron
exchanges. Hence, equation \eq{E:4.5} is the sum over all hard
rescattering correction amplitudes for pomeron exchange, in onium -
onium scattering. This approach is used in Refs. \cite{8,9,10}.  To
include Higgs production in the toy model, one has to replace one of
the $n$ dipole amplitudes, by the contribution $A_{H}(\delta\,Y_H)$ from the
subprocess for Higgs production. The leading subprocess, is the
quark triangle shown in \fig{tr}. Hence, for each of the terms, a
factor of $n$ is included, to account for the possibility that the
Higgs can be produced from any of the $n$ pomerons. After Higgs
production is included in the toy model of \eq{E:4.5}, the resulting
amplitude takes the form

\bea D_{Higgs}(Y\,,\,d)\,&=&
\,-\sum^{\infty}_{n=1}\Lb\,-1\Rb^n\,D_{Higgs}\Lb\,n\,,\,Y\,,\,d\Rb\,\,\notag\\
&=&\,\,-\sum^{\infty}_{n=1}(-1)^{n}d^{n-1}n!ne^{n\Delta\,Y}A_{H}\Lb\,\delta\,Y_H\Rb\,\,=\,\,\frac{\D}{\D\,d}\,D\Lb\,Y\,,\,d\Rb\,A_H\Lb\,\delta\,Y_H\Rb\,\notag\\
\label{E:4.6}\\
\notag\\
\mbox{where}\,\,\,\,\,\,\,\,\,\,\,\,\,\,D_{Higgs}\Lb\,n\,,\,Y\,,\,d\Rb\,&=&\,\,\,\,\,\,\,\,\,\,\,\,\,\,\,d^{n-1}\,n!\,n\,e^{n\,\,\Delta\,Y}\,A_H\Lb\,\delta\,Y_H\Rb\,\,\,\,\,\,\,\,\,\,\,\,=\,\,\frac{\D}{\D\,d}\,D\Lb\,n\,,\,Y\,,\,d\Rb\,A_H\Lb\,\delta\,Y_H\Rb\,\notag\\\label{E:4.6aa}
\eea

The notation $D_{Higgs}\Lb\,n\,,\,Y\,,\,d\Rb\,$ refers to the toy
model BFKL pomeron amplitude, including Higgs production, with $n$
pomeron branches. This should not be confused with the notation
$M_{Higgs}\Lb\,n\,,\,Y\Rb\,$, which refers to the equivalent
$n^{th}$ order term in the symmetric QCD dipole approach. The $n=1$
term in \eq{E:4.6}, corresponds to the single pomeron amplitude of
\fig{BFKL}. The $n=2$ term in equation \eq{E:4.6}, corresponds to
the first enhanced amplitude of \fig{BFKLenh}, with the hard
rescattering correction of the pomeron loop.  In section
\sec{sec:spdef}, the survival probability was defined by the
expression given in \eq{defsp}. Hence, in the toy model approach,
the survival probability takes the form

\beq
\,\,\frac{D_{Higgs}\Lb\,n=1\,,\,Y\,,\,d\Rb\,-\,\sum^{\infty}_{n=2}\Lb\,-1\Rb^nD_{Higgs}\Lb\,n\,,\,Y\,,\,d\Rb\,}{D_{Higgs}\Lb\,n=1\,,\,Y\,,\,d\Rb\,}\label{sptoy}
\eeq

Inspection of \eq{E:4.6} shows that the numerator on the RHS of
\eq{sptoy} can be rewritten as

\beq
D_{Higgs}\Lb\,Y\,,\,d\Rb\,=\,D_{Higgs}\Lb\,n=1\,,\,Y\,,\,d\Rb\,\,-\,\sum^{\infty}_{n=2}\Lb\,-1\Rb^nD_{Higgs}\Lb\,n\,,\,Y\,,\,d\Rb\,\label{sum}
\eeq

Hence, the toy model formula for the survival probability of \eq{sptoy} becomes

\beq
\,\,\frac{D_{Higgs}\Lb\,Y\,,\,d\Rb\,}{D_{Higgs}\Lb\,n=1\,,\,Y\,,\,d\Rb\,}\,=\,\frac{\frac{\D}{\D\,d}D\Lb\,Y\,,\,d\Rb\,}{\frac{\D}{\D\,d}D\Lb\,n=1\,,\,Y\,,\,d\Rb\,}\label{sptoy1}
\eeq

Typically, the Higgs signal will occupy a rapidity window
$\delta\,Y_H\,=\,\ln\frac{M_H^2}{4m^2}$
 Therefore, in the
toy model, pomeron exchange between scattering dipoles separated by
a rapidity gap of less than $\delta\,Y_H$, should be excluded for
Higgs production. Therefore, the toy model amplitude
$M_{Higgs}\Lb\,n\,,\,Y\,,\,d\Rb\,$ should be divided by the
scattering amplitude $M_{Higgs}\Lb\,n\,,\,\delta\,Y_H\,,\,d\Rb\,$,
which gives the scattering amplitude for dipoles separated by
a rapidity gap less than $\delta\,Y_H$. Taking this into account,
\eq{sptoy1} is
 modified to give the survival probability for diffractive Higgs
 production within the rapidity window $\delta\,Y_H$, as

\beq
<\vert\,S^2\vert\,>\,=\,\frac{\Lb\,\frac{\D}{\D\,d}D\Lb\,Y\,,\,d\Rb\,\Rb\,/\Lb\,\frac{\D}{\D\,d}D\Lb\,\delta\,Y_H\,,\,d\Rb\,\Rb\,}{\Lb\,\frac{\D}{\D\,d}D\Lb\,n=1\,,\,Y\,,\,d\Rb\,\Rb\,/\Lb\,\frac{\D}{\D\,d}D\Lb\,n=1\,,\,\delta\,Y_H\,,\,d\Rb\,\Rb\,}\label{sptoy11}
\eeq

In order to calculate the survival probability using the expression
of \eq{sptoy}, the value of the parameter $d$, appearing in the
expression for $D\Lb\,\,Y\,,\,d\Rb\,$, must be determined. To do so,
it is useful to refer back to the calculation of \sec{sec:sp2},
where the ratio
$\frac{M_{Higgs}\Lb\,n=2\,,\,Y\Rb\,}{M_{Higgs}\Lb\,n=1\,,\,Y\Rb\,}$
was calculated, in the symmetric QCD dipole approach (see
\eq{E:3.2.16}). In order for the toy model to be consistent with the
QCD dipole approach, the ratio calculated in \eq{E:3.2.16}, should
be the same in the toy model. Setting $n=1$, \eq{E:4.6aa} gives for
single pomeron amplitude in the toy model

\begin{equation}
D_{Higgs}(n=1,Y\,,\,d)=e^{\Delta\,Y}A_{H}\Lb\,\delta\,Y_H\Rb\,\label{E:4.7}
\end{equation}

 Setting $n=2$ in \eq{E:4.6aa}, the
first enhanced amplitude in the toy model is given by the following
expression

\begin{equation}
D_{Higgs}(n=2,Y\,,\,d)=-4de^{2\Delta\,Y}A_{H}\Lb\,\delta\,Y_H\Rb\,\label{E:4.8}
\end{equation}

Therefore,(using \eq{E:4.7} and \eq{E:4.8}), the following condition
is imposed

\beq
\frac{M_{Higgs}\Lb\,n=2\,,\,Y\Rb\,}{M_{Higgs}\Lb\,n=1\,,\,Y\Rb\,}\,=\,\frac{D_{Higgs}\Lb\,n=2\,,\,Y\,,\,d\Rb\,}{D_{Higgs}\Lb\,n=1\,,\,Y\,,\,d\Rb\,}\,=\,\frac{-4d^{2}e^{2\Delta\,Y}A_{H}\Lb\,\delta\,Y_H\Rb\,}{de^{\Delta\,Y}A_{H}\Lb\,\delta\,Y_H\Rb\,}=-4de^{\Delta\,Y}\label{condtn}
\eeq

Substituting for
$\frac{M_{Higgs}\Lb\,n=2\,,\,Y\Rb\,}{M_{Higgs}\Lb\,n=1\,,\,Y\Rb\,}$
the result of \eq{E:3.2.16} on the LHS of \eq{condtn}, and setting
the pomeron intercept equal to the BFKL intercept
$\Delta\,=\,\omega\Lb\,\nu\,=\,0\Rb\,$, to be consistent with the
QCD dipole approach, enables one to calculate a value for $d$ in the
toy model. One finds

\begin{equation}
d=0.7\label{E:4.10}
\end{equation}

One can now proceed to calculate the survival probability, by taking
into account all higher additional hard rescattering corrections,
using the formula of \eq{sptoy1}. From \eq{E:4.5}, the expression
for $D\Lb\,Y\,,\,d\Rb\,$ can be written as

\bea
D(Y\,,\,d)\,=\,-\sum^{\infty}_{n=1}n!\Lb\,-\,d\,e^{\Delta\,Y}\Rb^n\,&=&\,-\,\sum^{\infty}_{n=1}\,\int^{\infty}_{0}\,dt\,e^{-t}\Lb\,-\,\,d\,t\,e^{\Delta\,Y}\Rb^n\notag\\
&=&1\,-\,\int^{\infty}_{0}\,dt\,\frac{e^{-t}}{1+d\,t\,e^{\,\Delta\,Y}}\,\,\label{D11}
\eea After changing variables to
$\,\,\,u\,=\,\frac{1}{d\,\,e^{\,\Delta\,Y}}+t\,$, then the RHS
reduces to

\bea
D\Lb\,Y\,,\,d\Rb\,\,&=&1\,-\frac{\exp\Lb\,\frac{1}{de^{\,\Delta\,Y}}\Rb\,}{d\,e^{\,\Delta\,Y}}\,\int^{\infty}_{\frac{1}{d\,e^{\,\Delta\,Y}}}\,\,\frac{du\,e^{-\,u}}{u}\,\,=\,\,1\,-\frac{\exp\Lb\,\frac{1}{de^{\,\Delta\,Y}}\Rb\,}{d\,e^{\,\Delta\,Y}}\Gamma\Lb\,0\,,\,\frac{1}{d\,e^{\,\Delta\,Y}}\Rb\,\notag\\
\label{D111}
\eea

If one notes that in general
$\frac{d}{dx}\,\Gamma\Lb\,0\,,\,x\Rb\,\,=\,-\,\frac{e^{-x}}{x}$,
then substituting for $D\Lb\,Y\,,\,d\,\Rb\,$, the RHS of \eq{D111} in
the formula of \eq{sptoy11}, gives the following expression for the survival probability.

\beq
<\vert\,S^2\vert\,>\,=\,\Lb\,\frac{e^{2\,\Delta\,\delta\,Y_H}}{e^{2\,\Delta\,Y}}\Rb\,\Lb\,\frac{\exp\Lb\,\frac{1}{de^{\,\Delta\,Y}}\Rb\,\Gamma\Lb\,0\,,\,\frac{1}{d\,e^{\,\Delta\,Y}}\Rb\,\Lb\,\frac{1+d\,e^{\,\Delta\,Y}}{d\,e^{\,\Delta\,Y}}\Rb\,-d}{\exp\Lb\,\frac{1}{de^{\,\Delta\,\delta\,Y_H}}\Rb\,\Gamma\Lb\,0\,,\,\frac{1}{d\,e^{\,\Delta\,\delta\,Y_H}}\Rb\,\Lb\,\frac{1+d\,e^{\,\Delta\,\delta\,Y_H}}{d\,e^{\,\Delta\,\delta\,Y_H}}\Rb\,-d}\Rb\,\label{sptoy2}\eeq

The typical rapidity window $\delta\,Y_H$,  which the Higgs signal
is expected to occupy, is
$\delta\,Y_H\,=\,\ln\,\Lb\,\frac{M_H^2}{4m^2}\,\Rb\,$ where
$M_H^2\,\sim\,100GeV$. The typical rapidity gap is expected to be
$Y\,=\,19$ for
 the LHC energy of $\sqrt{s}\,=\,14\,TeV\,$.
 Setting the pomeron intercept equal to the
 BFKL intercept, $\Delta\,=\,\omega{}(\nu{}=0)=\bas\,4\ln{}2\approx\,0.34$ (see Ref. \cite{26}),
 the value for the survival probability from \eq{sptoy2}, is found to
be

\begin{equation}
<{}|S^{2}|{}>{}=0.004\label{E:4.17}
\end{equation}

This gives the survival probability as $0.4\%{}$. However, the
larger survival probability is obtained by abandoning the BFKL
intercept $\omega\Lb\,\nu\,=0\,\Rb\,\approx\,0.34\,$, and replacing the
intercept with that of the soft pomeron, $\Delta\,=\as\,=\,0.12$. In
this case, the RHS of \eq{sptoy2} using the soft pomeron intercept,
gives the following value for the survival probability,

\begin{equation}
<{}|S^{2}|{}>{}=0.23\label{E:4.15}
\end{equation}

Hence, using the value for the soft pomeron intercept
$\Delta\,=\as\,=\,0.12$, the survival probability is found to be
close to $22\%{}$. Alternatively, using a higher value for the
strong coupling $\as\,=\,0.25\,$, and replacing the intercept with
that of an upper limit for the soft pomeron intercept,
$\Delta\,=\,\as\,=\,0.25{}$, the survival probability from
\eq{sptoy2},  is found to be

\begin{equation}
<{}|S^{2}|{}>{}=0.022\label{E:4.16}
\end{equation}

Which is considerably lower and close to $2\%{}$. This value is
close to the value estimated by the Tel Aviv group in Ref. \cite{6},
and the Durham group in Ref. \cite{durham}. The values
found for the survival probability,
which depends on the choice of intercept are summarised below.

\begin{align}
&\mbox{{\bf Pomeron intercept $\Delta$}}&&\mbox{{\bf Survival probability $<\vert\,S^2\vert\,>$}}\notag\\
\notag\\
&\mbox{BFKL intercept $\,\,\,\,\,\,\,\,\,\,\,\,\,\,\,\,\,\Delta\,\,=\,\omega\Lb\,\nu\,=0\,\Rb\,\sim\,0.34$}&&\,0.004\notag\\
&\mbox{Soft pomeron intercept $\Delta\,=\,\as\,=\,0.25$}&&\,0.022\notag\\
&\mbox{Soft pomeron intercept $\Delta\,=\,\as\,=\,0.12$}&&\,0.23\notag
\end{align}
Therefore, from these results it is clear that the survival
probability depends critically on the intercept chosen. More specifically, the survival probability,
as a function of the intercept $\Delta$ is not monotonic.
The survival probability increases, as the intercept $\Delta$ decreases in value.
For large rapidity gaps $Y$, then from the formula of \eq{sptoy2}, the survival probability
is approximately proportional to

\beq
<\,\vert\,S^2\vert\,>\,\propto\,\frac{1}{\exp\Lb\,2\Delta\,\Lb\,\,Y\,-\,\delta\,Y_H\,\Rb\,\Rb\,}\label{Sprop}
\eeq
The typical LHC value for the rapidity gap $Y$ between scattering dipoles is $Y\,=\,19$, and for the predicted Higgs mass of $M_H^2$, the rapidity window occupied by the Higgs, is expected to be $\delta\,Y_H\,\,=\,\ln\Lb\,\frac{M_H^2}{4m^2}\Rb\,$. Hence, provided $Y\,-\,\delta\,Y_H\,>\,0$, then the expression of \eq{Sprop} explains why, the survival probability increases as the intercept $\Delta$ decreases.\\

Based on these results, in the toy model, the hard rescattering
contributions from higher $n$ corrections, range from $0.4\%{}$ up
to around $22\%{}$. Hence, the corrections are substantial and need
to be taken into account when calculating the survival probability.
$d$ in the toy model takes the value
 in \eq{E:4.10} $d=0.7$. This is less than unity. By inspection of the
summation in \eq{E:4.6}, one can see that $d$ is large enough, such
that the terms $n=3$ and higher, will give significant corrections
to the survival
probability calculated in this paper.\\

To summarise, it is found firstly that $d$ is large, giving
significant higher contributions. Secondly, these higher
contributions need to be taken into account, when calculating the
survival probability.

\section{Conclusion}
\label{sec:con}

The main results of this paper are the following.

\begin{enumerate}
\item \quad
The first calculation of the enhanced BFKL diagram for diffractive
Higgs production.

\item \quad
Estimates for the survival probability for the full set of enhanced
diagrams using the simplified toy model.

\item \quad
The results of this estimate for the survival probability, show that
the value depends crucially on the coupling constant of QCD, and that the
multi pomeron exchange gives a substantial contribution to the
survival probability.

\end{enumerate}

It was found that in the most consistent result for the survival
probability, the value is rather small, $0.4\%$. In conclusion, this
paper shows that hard processes give a substantial contribution in
the calculation of the survival probability. This paper is the first
step forward towards obtaining reliable estimates of the influences of hard
processes at high energy.

\section{Acknowledgements}
\label{sec:ack}

This paper is dedicated to my Mum and Dad. I would like to thank E.
Gotsman, A. Kormilitzin, E. Levin and A. Prygarin for fruitful
discussions on the subject. This research was supported in part  by
the Israel Science Foundation, founded by the Israeli Academy of
Science and Humanities, by a grant from the Israeli ministry of
science, culture \& sport, and the Russian  Foundation for Basic
research of the Russian Federation,   and by the BSF grant \
20004019.

\renewcommand{\thesection}{A-\arabic{section}}
\setcounter{section}{0}

\renewcommand{\thesubsection}{A-\arabic{subsection}}
\setcounter{subsection}{0}
\renewcommand{\theequation}{A-1-\arabic{equation}}
\setcounter{equation}{0}  

\renewcommand{\thesection}{A-\arabic{section}}
\setcounter{section}{0}
\appendix
\section{Appendix}
 \label{sec:app}

\renewcommand{\theequation}{A-1-\arabic{equation}}
\setcounter{equation}{0}  
\subsection{Calculation of the triple pomeron vertex}

In this section the triple pomeron vertex is calculated to give an
explicit expression in the momentum representation. This will be
useful for calculating the first enhanced diagram of \fig{BFKLenh}
in \sec{sec:BFKLenh}. In the expression for \fig{BFKLenh}, (see
\eq{E:3.1.7}), the BFKL functions $\omega\Lb\,\nu_1\Rb\,$ and
$\omega\Lb\,\nu_2\Rb\,$ are expanded around the saddle points
$\nu_1=\nu_2=0$. This gives the largest contribution to the
integration (see \eq{E:A19}). Hence, in this subsection the triple
pomeron vertex is calculated, in the limiting case when
$\nu_1=\nu_2=0$. It is assumed at the start of the calculation that
$\nu_1$ and $\nu_2$ are small and finite, however at the end of the
calculation $\nu_1$ and $\nu_2$ are put equal to zero. The triple
pomeron vertex shown in \fig{vpomqk} was defined in \sec{sec:3p}
(see \eq{E:trip}).

\bea
G_{3P}\Lb\vec{q},\vec{k},n=0,\gamma,\gamma_1,\gamma_2\Rb\,\,&=&\,\,G_{3P}\Lb\vec{q},\vec{k},\nu,\nu_1,\nu_2\Rb\,\,\label{E:A3a}\\
&=&\,\,\int{}\frac{d^2x_{10}d^2x_{20}d^2x_{30}}{x_{12}x_{23}x_{31}}E^{n,\nu}_{q}\Lb\,x_{10},x_{20}\Rb\,E^{n,\nu_1}_{k}\Lb\,x_{20},x_{30}\Rb\,E^{n,\nu_2}_{q-k}\Lb\,x_{30},x_{10}\Rb\nonumber
\eea

A useful expression, to be found in Ref. \cite{12}, was given in
\eq{E:3.1.8} in terms of the mixed representation of the vertex
function $E^{n,\nu}_{k}\Lb\,\vec{x}\Rb\,$ (see \eq{E:cc3}) as

\begin{equation}
G_{3P}(\vec{q}=0,\vec{k},\nu{},\nu_{1},\nu_{2})=\frac{1}{2\pi{}}\int\frac{d^{2}x_{01}}{x_{01}^{2}}x_{01}^{-2i\nu-1}e^{\frac{i\vec{k}\cdot\,\vec{x}_{01}}{2}}\int{}d^{2}x_{2}\frac{x_{01}^{2}}{x_{12}^{2}x_{02}^{2}}
x_{02}^{2i\nu_1+1}E_{k}^{n,\nu_1}(x_{02})x_{12}^{2i\nu_2+1}E_{-k}^{n,\nu_2}(x_{12})\label{E:A3}
\end{equation}

 In
\eq{E:A3}, it is assumed that $\vec{q}$ in \fig{vpomqk} is zero.
This is because for the calculation of the first enhanced diagram in
\sec{sec:BFKLenh} (see \fig{BFKLenh}), the momentum $\vec{q}$
transferred along the pomeron above and below the loop, is set to
zero, to make the calculation simpler.  In \fig{BFKLenh}, there are
two triple pomeron vertices, at opposite ends of the pomeron loop.
Here, the momentum $\vec{k}$ is the unknown momentum in the pomeron
loop.  Evaluating the integral over $x_{01}$ in \eq{E:A3} gives an
expression where the dependence on the momentum $\vec{k}$ is
explicit, namely \cite{12}

\bea G_{3P}(\vec{q}=0,\vec{k},\nu,\nu_{1},\nu_{2})&=&
2^{3-2\gamma-2\gamma_1-2\gamma_2}\Lb\,k^2\Rb\,^{i\nu+i\nu_1+i\nu_2-\frac{1}{2}}
\frac{\Gamma\Lb\,\frac{1}{2}-i\nu-i\nu_1-i\nu_2\Rb\,}
{\Gamma\Lb\,\frac{1}{2}+i\nu+i\nu_1+i\nu_2\Rb\,}
g_{3P}(\gamma,\gamma_1,\gamma_2)\,\,\,\,\,\,\,\,\,\,\,\,\,\label{E:A6}
\eea

where $g_{3P}\left(\gamma,\gamma_1,\gamma_2\right)$ is the
multidimensional integral related to the triple BFKL pomeron
interaction, given  by \cite{12}

\begin{equation}
g_{3P}(\gamma,\gamma_{1},\gamma_{2})=\int{}\frac{d^{2}x}{\vert{}x_{+}\vert{}^{2-2\gamma_{1}}\vert{}x_{-}\vert{}^{2-2\gamma_{2}}}\int{}\frac{d^{2}R}{\vert{}R_{+}\vert{}^{2\gamma_{1}}\vert{}R_{-}\vert{}^{2\gamma_{1}}}\int{}\frac{d^{2}R'}{\vert{}R'_{+}\vert{}^{2\gamma_{2}}\vert{}R'_{-}\vert{}^{2\gamma_{2}}}\vert{}R_{-}-R_{-}'\vert{}^{2\gamma{}+2\gamma_{1}+2\gamma_{2}-4}\label{E:A7}
\end{equation}

where in the notation of Ref. \cite{12},

\bea
&\gamma=\frac{1}{2}+i\nu&\,\,\,\,\,\gamma_1=\frac{1}{2}+i\nu_1\,\,\,\,\,\,\,\,\,\,\,\,\gamma_2=\frac{1}{2}+i\nu_2\notag\\
&x_{+}=x+\frac{n}{2}&\,\,\,\,\,x_{-}=x-\frac{n}{2}\,\,\,\,\,\,\,\,\,\,\,\,R_{+}=R+\frac{x_{+}}{2}\notag\\
&R_{-}=R-\frac{x_{+}}{2}&\,\,\,\,\,R'_{+}=R'+\frac{x_{-}}{2}\,\,\,\,\,R'_{-}=R'-\frac{x_{-}}{2}\label{E:A8}
\eea

Consider the part of the integration over $R'$ in \eq{E:A7}, which
takes the form

\bea\int\!\frac{d^2R'}{\Lb\,R_{+}'\Rb^{2\gamma_2}\Lb\,R_{-}'\Rb^{2\gamma_2}}\Lb\,R\_-R\_'\Rb^{2\gamma+2\gamma_1+2\gamma_2-4}
&=&\int\!\frac{dR'}{\Lb\,R_{+}'\Rb^{\gamma_2}\Lb\,R_{-}'\Rb^{\gamma_2}}\Lb\,R\_-R\_'\Rb^{\gamma+\gamma_1+\gamma_2-2}\notag\\
&\times\,&\int\!\frac{dR^{'\ast}}{\Lb\,R^{'\ast}_{+}\Rb^{\gamma_2}\Lb\,R^{'\ast}_{-}\Rb^{\gamma_2}}\Lb\,R^{\ast}\_-R^{'\ast}\_\Rb^{\gamma+\gamma_1+\gamma_2-2}\label{*}\eea

In \eq{*} the complex notation $d^2R'=dR'\,\,dR'^{\ast}$ has been
used. Evaluating the integrations over $R'$ and $R'^{\ast}$ gives

\bea\label{R'}
&\,&\int\!\frac{dR'}{\Lb\,R_{+}'\Rb^{\gamma_2}\Lb\,R_{-}'\Rb^{\gamma_2}}\Lb\,R\_-R\_'\Rb^{\gamma+\gamma_1+\gamma_2-2}\,\,\,=\notag\\
&\,&\,\,\,\pi^{\frac{1}{2}}\,\,\frac{\Gamma\Lb\,\frac{3}{2}-\gamma_1-\gamma_2-\gamma_1\Rb\,}{\Gamma\Lb\,2-\gamma-\gamma_1-\gamma_2\Rb\,}\frac{\Lb\,R_+-\Lb\,x_+-x\-\Rb\,\Rb^{\gamma_1+\gamma_2+\gamma-\frac{3}{2}}}{x\_^{\frac{1}{2}}}\,_2F_1\Lb\,\frac{1}{2},\frac{1}{2},\gamma_1+\gamma_2+\gamma-\frac{1}{2},\frac{R_{+}-\Lb\,x_{+}-x\-\Rb\,}{x\_}\Rb\,\nonumber\\
&+&\pi^{-\frac{1}{2}}\Gamma\Lb\,2-\gamma-\gamma_1-\gamma_2\Rb\,\Gamma\Lb\,\gamma+\gamma_1+\gamma_2-\frac{3}{2}\Rb\,x\_^{\Lb\,\gamma+\gamma_1+\gamma_2\Rb\,-2}\eea

Inspection of the right hand side of \eq{R'} shows that one has a
singularity at $\gamma=\gamma_1+\gamma_2+1$, in the limiting case
when $\nu_1=\nu_2=0$ (see \eq{E:A8}).  In this case,
$\Gamma\Lb\,2-\gamma-\gamma_1-\gamma_2\Rb\,$ tends to infinity,
which means that the first term on the RHS of \eq{R'} vanishes and
the second term gives the largest contribution. Therefore, in this
limiting case

\bea\label{Ra}
&\,&\int\!\frac{dR'}{\Lb\,R_{+}'\Rb^{\gamma_2}\Lb\,R_{-}'\Rb^{\gamma_2}}\Lb\,R\_-R\_'\Rb^{\gamma+\gamma_1+\gamma_2-2}\\
&=&\pi^{-\frac{1}{2}}\Gamma\Lb\,2-\gamma-\gamma_1-\gamma_2\Rb\,\Gamma\Lb\,\gamma+\gamma_1+\gamma_2-\frac{3}{2}\Rb\,x\_^{\Lb\,\gamma+\gamma_1+\gamma_2\Rb\,-2}\nonumber\eea

Inserting the result of \eq{Ra} back into the result of \eq{E:A7}
gives

\bea\label{3p}g_{3p}\Lb\,\gamma,\gamma_1,\gamma_2\Rb\,&=&\int\!\frac{dR_{+}}{R_{+}^{\gamma_2}}\int\!dx_{+}\frac{\Lb\,x_{+}-n\Rb^{\gamma+\gamma_1+\gamma_2-2}}{x_{+}^{\gamma_1}\Lb\,x_{+}-n\Rb\,^{\gamma_2}\Lb\,R_{+}-x_{+}\Rb^{\gamma_2}}\nonumber\\
&\times\,&\int\!\frac{d\Lb\,R_+^{\ast}\Rb\,}{\Lb\,R_+^{\ast}\Rb\,^{\gamma_2}}\int\!d\Lb\,x_+^{\ast}\Rb\,\frac{\Lb\,\Lb\,x_+^{\ast}\Rb\,-n\Rb^{\gamma+\gamma_1+\gamma_2-2}}{\Lb\,x_+^{\ast}\Rb\,^{\gamma_1}\Lb\,\Lb\,x_+^{\ast}\Rb\,-n\Rb\,^{\gamma_2}\Lb\,\Lb\,R_+^{\ast}\Rb\,-\Lb\,x_+^{\ast}\Rb\,\Rb^{\gamma_2}}\nonumber\\
&\times\,&\frac{1}{\pi}\Gamma^2\Lb\,2-\gamma-\gamma_1-\gamma_2\Rb\,\Gamma^2\Lb\,\gamma+\gamma_1+\gamma_2-\frac{3}{2}\Rb\,
\eea

Now, using the notation for $\gamma$, $\gamma_1$ and $\gamma_2$
defined in \eq{E:A8}, $g_{3p}\Lb\,\gamma,\gamma_1,\gamma_2\Rb\,$
becomes in the limit that $\nu_1=\nu_2=0$,

\bea\label{3p1}g_{3p}\Lb\,\gamma,\gamma_1,\gamma_2\Rb\,&=&\lim_{i\nu_1\rightarrow\,0}\int\!\frac{dR_{+}}{R_{+}^{\frac{1}{2}+i\nu_1}}\int\!dx_{+}\frac{1}{x_{+}^{\frac{1}{2}+i\nu_1}\Lb\,x_{+}-n\Rb\,^{\frac{1}{2}-3i\nu_1}\Lb\,R_{+}-x_{+}\Rb^{\frac{1}{2}+i\nu_1}}\nonumber\\
&\times\,&\lim_{i\nu_2\rightarrow\,0}\int\!\frac{dR^{\ast}_{+}}{\Lb\,R^{\ast}_{+}\Rb^{\frac{1}{2}+i\nu_2}}\int\!dx^{\ast}_{+}\frac{1}{\Lb\,x^{\ast}_{+}\Rb^{\frac{1}{2}+i\nu_2}\Lb\,x^{\ast}_{+}-n\Rb\,^{\frac{1}{2}-3i\nu_2}\Lb\,R^{\ast}_{+}-x^{\ast}_{+}\Rb^{\frac{1}{2}+i\nu_2}}\nonumber\\
&\times\,&\frac{1}{\pi}\Gamma^2\Lb\,\frac{1}{2}-i\nu\Rb\,\Gamma^2\Lb\,i\nu\Rb\,
\eea

It is instructive to leave $\nu_1$ and $\nu_2$ as small but finite
in the indices, and let them be driven to zero at the end of the
calculation, to avoid divergent integrals. Now integrating over
$x_{+}$ and $x_+^{\ast}$ gives the following result

\bea\label{3p2}&&g_{3p}\Lb\,\gamma,\gamma_1,\gamma_2\Rb\,=\,\frac{1}{\pi}\Gamma^2\Lb\,\frac{1}{2}-i\nu\Rb\,\Gamma^2\Lb\,i\nu\Rb\,\lim_{i\nu_1\rightarrow\,0}\int\!\frac{dR_{+}}{R_{+}^{1+2i\nu_1}}\times\,\\
&&\,\frac{\Gamma\Lb\,\frac{1}{2}+i\nu_1\Rb\,\Gamma\Lb\,\frac{1}{2}+3i\nu_1\Rb\,}{\Gamma\Lb\,1+4i\nu_1\Rb\,}\,_2F_1\Lb\,\frac{1}{2}+i\nu_1,\frac{1}{2}+i\nu_1,1+4i\nu_1,\frac{1}{R_{+}}\Rb\,+R_{+}^{1+2i\nu_1}\pi\,_2F_1\Lb\,\frac{1}{2}-3i\nu_1,\frac{1}{2}+i\nu_1,1,R_{+}\Rb\,\times\,\nonumber\\
&&\lim_{i\nu_2\rightarrow\,0}\int\!\frac{dR^{\ast}_{+}}{\Lb\,R^{\ast}_{+}\Rb^{1+2i\nu_2}}\times\,\notag\\
&&\frac{\Gamma\Lb\,\frac{1}{2}+i\nu_2\Rb\,\Gamma\Lb\,\frac{1}{2}+3i\nu_2\Rb\,}{\Gamma\Lb\,1+4i\nu_2\Rb\,}\,_2F_1\Lb\,\frac{1}{2}+i\nu_2,\frac{1}{2}+i\nu_2,1+4i\nu_2,\frac{1}{R^{\ast}_{+}}\Rb\,+\Lb\,R^{\ast}_{+}\Rb^{1+2i\nu_2}\pi\,_2F_1\Lb\,\frac{1}{2}-3i\nu_2,\frac{1}{2}+i\nu_2,1,R^{\ast}_{+}\Rb\,\notag
\eea

In the limit that $i\nu_1,i\nu_2\rightarrow\,0$ then the factor
$\frac{\Gamma\Lb\,\frac{1}{2}+i\nu_1\Rb\,\Gamma\Lb\,\frac{1}{2}+3i\nu_1\Rb\,}{\Gamma\Lb\,1+4i\nu_1\Rb\,}\rightarrow\,\pi$
and \eq{3p2} reduces to

\bea\label{3p2a}g_{3p}\Lb\,\gamma,\gamma_1,\gamma_2\Rb\,&=&\lim_{i\nu_1\rightarrow\,0}\int\!\frac{dR_{+}}{R_{+}^{1+2i\nu_1}}\Lb\,\,\,\,_2F_1\Lb\,\frac{1}{2},\frac{1}{2},1,\frac{1}{R_{+}}\Rb\,+R_{+}^{1+2i\nu_1}\,_2F_1\Lb\,\frac{1}{2},\frac{1}{2},1,R_{+}\Rb\,\Rb\,\nonumber\\
&\times\,&\lim_{i\nu_2\rightarrow\,0}\int\!\frac{dR^{\ast}_{+}}{\Lb\,R^{\ast}_{+}\Rb^{1+2i\nu_2}}\Lb\,\,\,\,_2F_1\Lb\,\frac{1}{2},\frac{1}{2},1,\frac{1}{R^{\ast}_{+}}\Rb\,+\Lb\,R^{\ast}_{+}\Rb^{1+2i\nu_2}\,_2F_1\Lb\,\frac{1}{2},\frac{1}{2},1,R^{\ast}_{+}\Rb\,\Rb\,\nonumber\\
&\times\,&\pi\Gamma^2\Lb\,\frac{1}{2}-i\nu\Rb\,\Gamma^2\Lb\,i\nu\Rb\,
\eea

Finally evaluating the integral over $R_{+}$ in \eq{3p2a} gives the
result for $g_{3p}\Lb\,\gamma,\gamma_1,\gamma_2\Rb$ as

\bea\label{3p3}g_{3p}\Lb\,\gamma,\gamma_1,\gamma_2\Rb\,&=&\,\frac{1}{4\nu_1\,4\nu_2\,\pi}\Gamma^2\Lb\,\frac{1}{2}-i\nu\Rb\,\Gamma^2\Lb\,i\nu\Rb\,
\eea

Substituting this result for
$g_{3P}\Lb\,\gamma,\gamma_1,\gamma_2\Rb\,$ of \eq{3p3} into the
expression of \eq{E:A6}, the triple pomeron vertex is given
explicitly in the momentum representation, in the limit that
$\nu_1=\nu_2=0$, by the expression,

\bea
G_{3P}(\vec{q}=0,\vec{k},\nu,\nu_{1}\rightarrow\,0,\nu_{2}\rightarrow\,0)\,\,=\,\,\frac{2^{-2i\nu}}{4\nu_{1}4\nu_{2}\pi}\Lb\,k^2\Rb\,^{i\nu-\frac{1}{2}}\frac{\Gamma^3\Lb\,\frac{1}{2}-i\nu\Rb\,\Gamma^2\Lb\,i\nu\Rb\,}{\Gamma\Lb\,\frac{1}{2}+i\nu\Rb\,}\label{E:A12}
\eea

To calculate the first enhanced amplitude of \fig{BFKLenh}, there is
an integration to be evaluated, of the two triple pomeron vertices
at both ends of the loop, over the unknown momentum $\vec{k}$ (see
\eq{E:3.1.7}), which takes the form

\beq
\int{}d^{2}k\,\,G_{3P}(\vec{q}=0,\vec{k},\nu,\nu_{1},\nu_{2})G_{3P}(\vec{q}=0,\vec{k},-\nu{}',\nu_{1},\nu_{2})
\eeq

Inserting the result of \eq{E:A12} gives

\begin{align}
&\int{}d^{2}k\,\,\,G_{3P}(\vec{q}=0,\vec{k},\nu,\nu_{1},\nu_{2})G_{3P}(-\nu{}',\nu_{1},\nu_{2})
=\notag\\
&\frac{2^{2\Lb\,i\nu'-i\nu\Rb\,}}{\pi^2}\int{}\textit{d}^{2}
\textit{k}\Lb\,k^2\Rb\,^{(i\nu{}-i\nu{}')-1}
\frac{\Gamma^3\Lb\,\frac{1}{2}-i\nu{}\Rb\,\Gamma^3\Lb\,\frac{1}{2}+i\nu{}'\Rb\,\Gamma^2\Lb\,i\nu\Rb\,\Gamma^2\Lb\,-i\nu'\Rb\,}
{\Gamma\Lb\,\frac{1}{2}+i\nu{}\Rb\,\Gamma\Lb\,\frac{1}{2}-i\nu{}'\Rb\,}
\frac{1}{16\nu^2_{1}16\nu^2_{2}} \label{E:12a}
\end{align}

Now to integrate over $k$, it is useful to make the change of
variable $l=\ln{}k$. Then the right hand side of \eq{E:12a} reduces
to a delta function in $\nu$ and $\nu'$, to give the result

\bea
&\,\,&\int{}d^{2}k\,\,\,G_{3P}(\vec{q}=0,\vec{k},\nu,\nu_{1},\nu_{2})G_{3P}(-\nu{}',\nu_{1},\nu_{2})\notag\\
&=&
\frac{2^{2\Lb\,i\nu'-i\nu\Rb\,}}{8}\frac{\delta\Lb\,\nu-\nu'\Rb\,}{4\nu^2_{1}4\nu^2_{2}}
\frac{\Gamma^3\Lb\,\frac{1}{2}-i\nu{}\Rb\,\Gamma^3\Lb\,\frac{1}{2}+i\nu{}'\Rb\,\Gamma^2\Lb\,i\nu\Rb\,\Gamma^2\Lb\,-i\nu'\Rb\,}
{\Gamma\Lb\,\frac{1}{2}+i\nu{}\Rb\,\Gamma\Lb\,\frac{1}{2}-i\nu{}'\Rb\,}
\label{E:A13} \eea

\renewcommand{\theequation}{A-2-\arabic{equation}}
\setcounter{equation}{0}  
\subsection{Calculation of the first enhanced amplitude}

Once the integral over the unknown momentum $\vec{k}$ in the pomeron
loop in \fig{BFKLenh} has been evaluated,  the first enhanced
diagram with Higgs production can be calculated from \eq{E:3.1.7}.
Inserting the result of \eq{E:A13} in the right hand side of
equation \eq{E:3.1.7}, gives

\begin{align}
P^{BFKL
}_{enhanced}=&\,\,\,\,\frac{B\pi^2}{8}\int\!d\nu\,d\nu_{1}d\nu_{2}\int^{Y}_{Y_H  +  \h \delta Y_H } dY_{1}\int^{Y_{H} - \h \delta Y_H}_{0}dY_{2}\notag\\
&\times\,E\Lb\,\vec{p}_1,\vec{q},\nu{}\Rb\,\mathcal{D}^{2}(\nu)e^{\omega(\nu)(Y-Y_{1}+Y_{2})}\frac{\Gamma{}^{2}\Lb\,\frac{1}{2}-i\nu{}\Rb\,\Gamma{}^{2}\Lb\,\frac{1}{2}+i\nu{}\Rb\,}{4\nu_{1}^{2}4\nu_{2}^{2}\nu{}^{2}\sin{}^{2}\left(i\nu{}\pi{}\right)}\notag\\
&\times\mathcal{D}(\nu_{1})\mathcal{D}(\nu_{2})e^{(\omega(\nu_{1})+\omega(\nu_{2}))(Y_{1}-Y_{2})-\omega\Lb\,\nu_2\Rb\,\ln\Lb\,\frac{M_H^2}{4m^2}\Rb\,}E\Lb\,\vec{p}_2,\vec{q},-\nu{}\Rb\,\,\label{E:A14a}\\
\mbox{where}\,\,\,\,\,B=&\,\,\,\,-\frac{\alpha{}^{4}_{s}\pi{}^{4}}{8}\left(\frac{\alpha_{s}N_{C}}{2\pi{}^{2}}\right)^{2}\notag
\end{align}

The largest contribution to the integral over $\nu$ in \eq{E:A14a},
is when $i\nu\rightarrow\,\frac{1}{2}$, because $\omega{}(\nu{})$
has a pole at $i\nu\rightarrow\frac{1}{2}$, such that \beq\label{Oh}
\omega{}(\nu{})\xrightarrow{i\nu\rightarrow\frac{1}{2}}\frac{2\bar{\alpha{}_{s}}}{i\nu{}-\frac{1}{2}}
\eeq

It is assumed, that the conjugate momenta $\vec{p}_1$ and
$\vec{p}_2$ of the two scattering dipoles in \fig{BFKLenh}, are
equal. Using the expression of \eq{E:Cvert} for the two pomeron
vertices, \eq{E:A14a} reduces to

\begin{align}
P^{BFKL}_{enhanced}=&-\frac{2\,B\pi^6}{p^2}\int\!d\nu\,d\nu_{1}d\nu_{2}\int^{Y}_{Y_H + \delta Y_H}dY_{1}\int^{Y_H - \h \delta Y_H}_{0}dY_{2}\frac{\exp\left(\frac{2\bar{\alpha{}_{s}}(Y-Y_{1}+Y_{2})}{\left(\frac{1}{2}-i\nu{}\right)}\right)}{\Lb\,\frac{1}{2}-i\nu\Rb^6}\notag\\
&\times\,\frac{\mathcal{D}(\nu_{1})\mathcal{D}(\nu_{2})}{4\nu_{1}^{2}4\nu_{2}^{2}}\,e^{(\omega(\nu_{1})+\omega(\nu_{2}))(Y_{1}-Y_{2})-\omega\Lb\,\nu_2\Rb\,\ln\Lb\,\frac{M_H^2}{4m^2}\Rb\,}\,\label{E:A14}
\end{align}

One has the singularity as $i\nu\rightarrow\frac{1}{2}$, in the
integrand of \eq{E:A14}, due to the factor
$\frac{1}{\Lb\,i\nu-\frac{1}{2}\Rb^6}$. To remove this singularity,
the substitution \beq
\frac{\exp\left(\frac{\bar{\alpha{}}_{s}(Y-Y_{1}+y_{2})}{\frac{1}{2}-i\nu{}}\right)}{\left(i\nu-\frac{1}{2}\right)^{6}}=
\frac{1}{\Lb\,2\bar{\alpha{}}_{s}\Rb^{4}}\frac{d^{4}}{dY^{4}}\left(\frac{1}{\left(i\nu-\frac{1}{2}\right)^{2}}\exp\left(\frac{\bar{2\,\alpha}_{s}(Y-Y_{1}+Y_{2})}{\frac{1}{2}-i\nu{}}\right)\right)\label{trick}\eeq
is used. The integrand in \eq{E:A14} then simplifies to

\begin{align}
P^{BFKL}_{enhanced}=&-\frac{2B\pi^6}{p^2}\frac{1}{\Lb\,2\bas\Rb^4}\frac{d^4}{dY^4}\int\!d\nu\,d\nu_{1}d\nu_{2}\int^{Y}_{Y_H + \delta Y_H}dY_{1}\int^{Y_H  - \delta Y_H}_{0}dY_{2}\frac{\exp\left(\frac{2\bar{\alpha{}_{s}}(Y-Y_{1}+Y_{2})}{\left(\frac{1}{2}-i\nu{}\right)}\right)}{\Lb\,\frac{1}{2}-i\nu\Rb^2}\notag\\
&\times\,\frac{\mathcal{D}(\nu_{1})\mathcal{D}(\nu_{2})}{4\nu_{1}^{2}4\nu_{2}^{2}}\,e^{(\omega(\nu_{1})+\omega(\nu_{2}))(Y_{1}-Y_{2})-\omega\Lb\,\nu_2\Rb\,\ln\Lb\,\frac{M_H^2}{4m^2}\Rb\,}\,\label{E:A16}
\end{align}

There is still the singularity in the integrand due to the factor
$\frac{1}{\Lb\,\frac{1}{2}-i\nu\Rb^2}$. To remove this singularity,
it is useful to change the variables such that

\beq\,u=\frac{2\bas}{\Lb\,\frac{1}{2}-i\nu\Rb}\eeq

Then \eq{E:A16} simplifies to

\begin{align}
P^{BFKL}_{enhanced}=&-\frac{i\,B\pi^6}{p^2}\frac{1}{\Lb\,2\bas\Rb^5}\frac{d^4}{dY^4}\int\!du\,d\nu_{1}d\nu_{2}\int^{Y}_{Y_H + \h \delta Y_H}dY_{1}\int^{Y_{H}- \h \delta Y_H}_{0}dY_{2}\exp\Lb\,\Lb\,Y-Y_{1}+Y_{2}\Rb\,u\Rb\,\notag\\
&\times\,\frac{\mathcal{D}(\nu_{1})\mathcal{D}(\nu_{2})}{4\nu_{1}^{2}4\nu_{2}^{2}}\,e^{(\omega(\nu_{1})+\omega(\nu_{2}))(Y_{1}-Y_{2})-\omega\Lb\,\nu_2\Rb\,\ln\Lb\,\frac{M_H^2}{4m^2}\Rb\,}\,\label{E:A17}
\end{align}

Integrating over $u$,  the right hand side is found to be
proportional to a delta function in the rapidity,

\begin{align}
P^{BFKL}_{enhanced}=&\frac{2B\pi^7}{p^2}\frac{1}{\Lb\,2\bas\Rb^5}\frac{d^4}{dY^4}\int\!d\nu_{1}d\nu_{2}\int^{Y}_{Y_H +\h \delta Y_H}dY_{1}\int^{Y_{H}- \h \delta Y_H}_{0}dY_{2}\,\delta\Lb\,Y-Y_1+Y_2\Rb,\notag\\
&\times\,\frac{\mathcal{D}(\nu_{1})\mathcal{D}(\nu_{2})}{4\nu_{1}^{2}4\nu_{2}^{2}}\,e^{(\omega(\nu_{1})+\omega(\nu_{2}))(Y_{1}-Y_{2})-\omega\Lb\,\nu_2\Rb\,\ln\Lb\,\frac{M_H^2}{4m^2}\Rb\,}\,\label{E:A18}
\end{align}

The integration over the two rapidity variables $Y_1$ and $Y_2$ is
now simple, because of the delta function in the integrand. The BFKL
functions $\omega\Lb\nu_{1}\Rb$ and $\omega\Lb\nu_{2}\Rb$, can be
expanded around the saddle points $\nu_{1}=0$ and $\nu_{2}=0$. From
the definition of \eq{E:3.1.3}, $\mathcal{D}\Lb\,\nu_1\Rb\,$ and
$\mathcal{D}\Lb\,\nu_2\Rb\,$ would vanish at $\nu_1=0$ and
$\nu_2=0$. However, the factors of $\nu_1^2$ and $\nu_2^2$ are
canceled by the $\nu_1^2$ and $\nu_2^2$, appearing in the
denominator in the integrand of \eq{E:A18}. Hence, the integrals
over $\nu_1$ and $\nu_2$ can be evaluated at the typical values
$\nu_1=0$ and $\nu_2=0$, to give the result

\bea
P^{BFKL}_{enhanced}\,&\,=\,&\,\frac{32B\pi^7}{\Lb\,2\bas\Rb^5p^2}\,\frac{d^{4}}{dY^{4}}\int{}d\nu_{1}d\nu_{2}\h\,\delta\,Y_H\notag\\
&\times\,&\exp\left(\omega\Lb\,\nu_1\,=0\Rb\,Y+\omega\Lb\,\nu_2\,=0\,\Rb\,Y+\frac{1}{2}\nu_{1}^2\omega\,"(\nu_1\,=\,0\,)Y+\frac{1}{2}\nu_{2}^2\omega\,"(\nu_2\,=\,0)Y\right)\,\notag\\
&\times\,&e^{-\omega\Lb\,\nu_2\Rb\,\ln\Lb\,\frac{M_H^2}{4m^2}\Rb\,}\label{E:A19}
\eea

Finally, there are the integrations  over $\nu_1$ and $\nu_2$ to
evaluate, which are just Gaussian integrals. Therefore, the result
for the right hand side of \eq{E:A19} is

\bea
P^{BFKL}_{enhanced}&=&\frac{32\,B\pi^8\,}{\Lb\,2\bas\Rb^5p^2}
\frac{\delta\,Y_H}{\omega"\Lb\,\nu\,=0\,\Rb\,}
\Lb\,\frac{\Lb\,2\omega^2\Lb\,\nu\,=0\,\Rb\,\Rb^4\,}{Y}
-4\,\frac{\Lb\,2\omega\Lb\,\nu=0\Rb\,\Rb^3}{Y^2}
+12\frac{4\omega^2\Lb\,\nu=0\Rb\,}{Y^3}-12\frac{2\omega\Lb\,\nu=0\Rb\,}
{Y^4}+\frac{4}{Y^5}\Rb\,\notag\\
&\times\,&\,e^{2\omega\Lb\,\nu\,=0\,\Rb\,\Lb\,Y-\h\,\ln\Lb\,\frac{M_H^2}{4m^2}\Rb\,\Rb\,}\,
\label{E:A20int}
\eea

Taking the rapidity $Y$ to be $19$ for the LHC energy $\sqrt{s}\,=\,14000\,GeV\,$, and $\omega\Lb\,\nu\,=0\,\Rb\,=\,4\bas\,\ln\,2$, then the first and second terms in the brackets of \eq{E:A20int} are the largest terms, and hence at leading order

\beq
P^{BFKL}_{enhanced}\,=\,\frac{32\,B\pi^8\,}{\Lb\,2\bas\Rb^5p^2}\frac{\delta\,Y_H}{\omega"\Lb\,\nu\,=0\,\Rb\,}\Lb\,\frac{\Lb\,2\omega\Lb\,\nu\,=0\,\Rb\,\Rb^4\,}{Y}-4\frac{\Lb\,2\omega"\Lb\,\nu\,=0\,\Rb\,\Rb^3}{Y^2}\Rb\,\,e^{2\omega\Lb\,\nu\,=0\,\Rb\,\Lb\,Y-\h\,\ln\Lb\,\frac{M_H^2}{4m^2}\Rb\,\Rb\,}\,\label{E:A20}
\eeq

\bibliographystyle{amsplain}
\bibliography{}
\label{sec:bib}

\end{document}